\documentclass{acmart}
\usepackage{multirow}
\usepackage{hyperref}
\usepackage{subcaption}
\usepackage{graphicx}
\usepackage{caption}
\usepackage[algo2e,linesnumbered,ruled,vlined,boxed]{algorithm2e} 
\AtBeginDocument{%
  }

\setcopyright{acmlicensed}
\copyrightyear{2018}
\acmYear{2018}
\acmDOI{XXXXXXX.XXXXXXX}

\acmConference[Conference acronym 'XX]{Make sure to enter the correct
  conference title from your rights confirmation emai}{June 03--05,
  2018}{Woodstock, NY}
\acmISBN{978-1-4503-XXXX-X/18/06}
\newcommand{\Prob}[1]{{\textbf{Pr}} \left( #1 \right)}
\newcommand{\Var}[1]{{\textbf{Var}} \left[ #1 \right]}

\newcommand{\Expec}[1]{\textbf{E} \left[ #1 \right]}
\newcommand{\CExpec}[2]{\textbf{E}_{#2}\left[ #1\right]}

\newcommand{\prk}{\texttt{p-rk}}
\newcommand{\pab}{\texttt{p-ab}}

\newcommand{\calS}{\mathcal{S}}

\newcommand{\bigO}{\mathcal{O}}

\newcommand{\calF}{\mathcal{F}}

\newcommand{\calD}{\mathcal{D}}

\newtheorem{definition}{Definition}

\newlength{\commentWidth}
\let\oldnl\nl
\newcommand{\nonl}{\renewcommand{\nl}{\let\nl\oldnl}}
\DontPrintSemicolon
\SetKwInOut{Input}{Input}\SetKwInOut{Output}{Output}
\usepackage{algorithmic}

\newcommand{\argmax}{\texttt{argmax}}

\usepackage{xspace}
\usepackage{booktabs}

\usepackage{hyperref}
\usepackage{mathtools}
\usepackage{amsmath}
\usepackage{amsfonts}
\usepackage{bm}
\usepackage{dsfont}
\usepackage{upgreek}

\begin{document}

\title{Fast Estimation of Percolation Centrality}

\author{Antonio Cruciani}
\authornote{}
\email{antonio.cruciani@gssi.it}
\orcid{}
\affiliation{%
	\institution{Gran Sasso Science Institute}
	\city{}
	\state{}
	\country{Italy}
}

\renewcommand{\shortauthors}{Antonio Cruciani}

\begin{abstract}
	In this work we present a new algorithm to approximate the percolation centrality of every node in a graph. Such a centrality measure quantifies the importance of the vertices in a network during a contagious process. In this paper, we present a randomized approximation algorithm that can compute probabilistically guaranteed high-quality percolation centrality estimates, generalizing techniques used by Pellegrina and Vandin (TKDD 2024) for the betweenness centrality. The estimation obtained by our algorithm is within $\varepsilon$ of the value with probability at least $1-\delta$, for fixed constants $\varepsilon,\delta \in (0,1)$. We support
	our theoretical results with an extensive experimental analysis on several	real-world networks and provide empirical evidence that our algorithm improves the current state of the art in speed, and sample size while maintaining high accuracy of the percolation centrality estimates. 
\end{abstract}



\received{20 February 2007}
\received[revised]{12 March 2009}
\received[accepted]{5 June 2009}

\maketitle
\section{Introduction}
Finding the most important nodes in a graph is a well known task in graph mining. A wide spread technique to assign an importance score to the nodes is to use a centrality measure. In this paper we consider the \emph{percolation centrality}, a measure used in setting where graphs are used to model a contagious process in a network (e.g., infection or misinformation spreading). The percolation centrality is a generalization of the betweenness centrality that is defined as the fraction of shortest paths passing through a node $v$ over the overall number of shortest paths in a graph. The percolation centrality introduces weights on the shortest paths, and the weight of a shortest path depends on the disparity between the level of contamination of the two end vertices of such path. The percolation centrality has been introduced by~\cite{Broadbent_1957} to model the passage of a fluid in a medium. Subsequently, such centrality measure was adapted to graphs by Piraveenan et al.~\cite{Piraveenan_2013}, in which the medium  are the vertices of a given graph $G$ and each vertex $v$ in $G$ has a percolation state that reflects the level of contamination of the node $v$. The best technique to compute the exact percolation centrality scores of all the nodes in a graph $G$ is to solve the All Pair Shortest Paths (APSP) problem (i.e., to run a Breath First Search or Dijkstra algorithm from each node $v$). Unfortunately, under the APSP conjecture~\cite{Abboud_2014,Abboud_2015} no truly subcubic time ($\bigO(n^{3-\varepsilon})$ for $\varepsilon >0$) combinatorial algorithm can be designed. Lima et al.~\cite{Lima_2020,Lima_2022} used techniques proposed by Riondato and Kornaropoulos~\cite{Riondato_2016} and Riondato and Upfal~\cite{Riondato_2018} for the betweenness centrality to design efficient randomized algorithms for approximating the percolation centrality. In both cases, the authors provided algorithms that sample a subset of all shortest paths in the graph so that, for a given $\varepsilon,\delta\in (0,1)$ they obtain values within $\varepsilon$ from the exact value with probability $1-\delta$. The techniques used in~\cite{Lima_2020} relies on pseudo-dimension (a generalization of the VC-Dimension) theory to provide a \emph{fixed sample size} approximation algorithm 
to compute an approximation of the percolation centrality of all the nodes. More precisely, Lima et al.~\cite{Lima_2020} proved that when $\bigO(\log (VD(G)/\delta)/\varepsilon^2) $ percolated shortest paths are sampled uniformly at random, the approximations are within an additive error $\varepsilon$ of the exact centralities with probability at least $1-\delta$, where $VD(G)$ is the vertex diameter\footnote{The vertex diameter of a graph is the number of nodes in the longest shortest path. On unweighted graphs, the vertex diameter $VD(G) = D(G) +1$ where $D(G)$ is the diameter of the graph $G$.} of the graph. 
Next, Lima et al.~\cite{Lima_2022} combined their previous results on pseudo-dimension with the ones proposed by Riondato and Upfal in~\cite{Riondato_2018} on Rademacher Averages applied to the betweenness centrality. They showed that this combination can be further developed for giving an approximation algorithm for the percolation
centrality based on a \emph{progressive sampling} strategy. Their algorithm iteratively increases the sample size until the desired accuracy is achieved. The stopping condition depends on the Rademacher Averages of the current sample of shortest paths.

\paragraph{Our Contribution.} In this work we generalize the techniques developed in the work of Pellegrina and Vandin~\cite{Pellegrina_2024} on the betweenness centrality to the percolation centrality. More precisely

\begin{itemize}
    \item We generalize the Balanced Bidirectional Breadth First Search algorithm used to sample shortest paths for the betweenness centrality~\cite{Borassi_2019,Pellegrina_2024} to the computation of percolated shortest paths. 
    
    \item We derive a new bound on the sufficient number of samples to approximate the
    percolation centrality for all nodes, that is governed by the sum of the percolation centrality of all the nodes and the maximum variance of the percolation centrality estimators instead of the vertex diameter in~\cite{Lima_2020}.
    Moreover, this result solves an open problem in~\cite{Pellegrina_2023,Pellegrina_2024} on whether the sample complexity bounds for the betweenness can be efficiently extended to the percolation centrality. As a consequence, it significantly improves on the state-of-the-art results for the percolation centrality estimation process.
  
    \item We define a progressive sampling algorithm that uses an
    advanced tool from statistical learning theory, namely \emph{Monte Carlo Empirical
    Rademacher Averages}~\cite{Barlett_2003} and the above results to provide a high quality approximation of the percolation centrality. The algorithm's output is a function of two parameters $\varepsilon\in (0,1)$ controlling the approximation's accuracy and $\delta\in (0,1)$ controlling the confidence of the computed approximation. 

	\item  We perform an extensive experimental evaluation showing that our algorithm improves on the state-of-the-art in terms of running time and sample size.
\end{itemize}

\section{Preliminaries}
We now introduce the definitions, notation and results we use as the
groundwork of our proposed algorithm. 

\subsection{Graphs and Percolation centrality}
Given a graph $G= (V,E)$ the percolation states $x_v$ for each $v\in V$ and couple of nodes $(s,z)\in V\times V$, we define $\Gamma_{sz}$ to be the set of all the shortest paths from $s$ to $z$, and $\sigma_{sz} = |\Gamma_{sz}|$. For a given path $p_{sz}\in\Gamma_{sz}$, we define $\texttt{Int}(p_{sz})$ be the set of internal vertices of $p_{sz}$, i.e., $\texttt{Int}(p_{sz}) = \{v\in V : v\in p_{sz}\land v\neq s\neq z\}$. Moreover, we denote $\sigma_{sz}(v)$ as the number of shortest paths from $s$ to $z$ that passes through $v$. Let $0\leq x_v\leq 1$ be the percolation state of $v\in V$. We say $v$ is \emph{fully percolated} if $x_v =1$, \emph{non-percolated} if $x_v= 0$ and \emph{partially percolated} if $0<x_v<1$. Moreover, we say that a path from $s$ to $z$ is percolated if $x_s-x_z>0$. The percolation centrality is defined as follows.
\begin{definition}[Percolation Centrality] Let $R(x) = \max\{0,x\}$ be the \emph{ramp function}. Given a graph $G = (V,E)$ and percolation states $x_v$ for each $v\in V$, the percolation centrality of a node $v\in V$ is defined as
	\begin{align*}
		p(v)=\frac{1}{n(n-1)}\sum_{\substack{(s,z)\in V\times V\\ s\neq v\neq z}}{\frac{\sigma_{sz}(v)}{\sigma_{sz}}\cdot \kappa(s,z,v)}
	\end{align*}
 Where $\kappa(s,z,v) = \frac{R(x_s-x_z)}{\sum_{\substack{(u,w)\in V\times V\\ u\neq v\neq w}}R(x_u-x_w)}$
\end{definition}
Finally, we define the \emph{average number of internal nodes} in a shortest path as
\begin{align*}
    \rho = \frac{1}{n(n-1)} \sum_{s,z\in V}|\texttt{Int}(p_{sz})|
\end{align*}

\subsection{Percolation Centrality Estimators}
We review existing approaches for the percolation centrality estimation. More precisely, Lima et al.~\cite{Lima_2020,Lima_2022} proposed two estimators to approximate the percolation centrality. For each of the two approaches we define a domain $\calD$, a family $\calF$ of functions from $\calD$ to $[0,1]$, and a probability distribution $\pi$ over $\calD$.
\paragraph{The Riondato Kornaropoulos based.} Lima et al.~\cite{Lima_2020}, introduced a first estimator based on the Riondato and Kornaropoulos one~\cite{Riondato_2016} for the betweenness centrality that is tailored to their use of VC-dimension for the analysis of the sample size sufficient to obtain a high-quality estimate of the percolation centrality of all nodes. We refer to such estimator as the Percolation based on the Riondato and Kornapolus estimator (\prk~for short). The domain $\calD_{\prk}$ is the set of all shortest paths between all pairs of vertices in the graph $G$, i.e.,
\begin{align*}
	\calD_{\prk} = \bigcup_{\substack{(s,z)\in V\times V\\s\neq z}} \Gamma_{sz}
\end{align*}
Let $p_{sz}\in \calD_{\prk}$ be any shortest path from $s$ to $z$. The distribution $\pi_{\prk}$ over $\calD_{\prk}$ assigns to $p_{sz}$ the probability mass 
\begin{align*}
	\pi_{\prk}(p_{sz}) = \frac{1}{n(n-1)\sigma_{sz}}
\end{align*}
Lima et al., extended the efficient sampling scheme for the betweenness centrality~\cite{Riondato_2018} to draw independent samples from $\calD_{\prk}$ according to $\pi_{\prk}$. The family $\calF_{\prk}$ of functions contains one function $f_v:\calD_{\prk}\to [0,1]$ for each vertex $v$, defined as
\begin{align*}
	f_v(p_{sz}) = \left\{\begin{array}{cl}
		\kappa(s,z,v) &\text{if }v\in \texttt{Int}(p_{sz})\\
		0 &\text{otherwise}
	\end{array}\right.
\end{align*} 
\paragraph{The ABRA based.}
Subsequently, Lima et al.~\cite{Lima_2022} proposed another estimator based on the ABRA estimator~\cite{Riondato_2018} (we refer to such estimator as \pab) over the domain 
\begin{align*}
	\calD_{\pab} = \{(s,z)\in V\times V: s\neq z\}
\end{align*}
with uniform distribution $\pi_{\pab}$ over $\calD_{\pab}$ and family $\calF_{\pab}$ containing a function $f_v$ for each vertex $v$ defined as follows
\begin{align*}
	f_v(s,z) = \frac{\sigma_{sz}(v)}{\sigma_{sz}} \cdot \kappa(s,z,v)\in [0,1]
\end{align*}
\paragraph{Statistical Properties of the estimators.} We might ask ourselves, which of the two reviewed percolation centrality estimators provide the best estimates (i.e., should we use). To answer this question, we observe that both of them are ``built on top'' of some (e.g.~\cite{Riondato_2016,Riondato_2018}) specific betweenness estimator. Furthermore, such a question has already been answered for the betweenness centrality by Cousins et al., in~\cite{Cousins_2023}. This inherent relation between the percolation and the betweenness centrality makes it easy to extend the analysis of the work in~\cite{Cousins_2023} to the percolation centrality estimators (\texttt{p-rk} and \texttt{p-ab}).
Indeed, the key observation is that the two estimators are equal (to the exact percolation centrality) in
expectation, and each simply commutes progressively less randomness from the inner sample to
the outer expectations. Each sampling algorithm may therefore be seen as a progressively more
random stochastic approximation of the exact algorithm. Intuitively, when we use the estimators with $r = 1$ samples, their relationship can be expressed as
\begin{align*}
    p(v) = \overbrace{\CExpec{\underbrace{\CExpec{f_v(p_{sz})}{p_{sz}\in\Gamma_{sz}}}_{\texttt{p-rk}}}{s\neq z}}^{\texttt{p-ab}}
\end{align*}
Where $s$ and $z$ ($s\neq z$) are sampled uniformly at random from $V$ and $p_{sz}$ is sampled uniformly at random from $\Gamma_{sz}$. In other words, each estimator computes a conditional expectation of the percolation centrality. \texttt{p-ab} conditions on the vertices $s$ and $z$, and \texttt{p-rk} conditions on a randomly sampled percolated shortest path $p_{sz}\in\Gamma_{sz}$. Furthermore, by using the results for the variance of the betweenness estimators~\cite{Cousins_2023}, we naturally obtain the following corollary for the percolation centrality estimators.
\begin{corollary}[Of Lemma 4.5 in~\cite{Cousins_2023}]\label{corollary:variance}
Given a graph $G = (V,E)$, for every $v\in V$ it holds:
    $$\frac{\Var{f_v(s,z)}}{\Var{f_v(p_{sz})}}\leq \max_{\substack{s,z \in V\\ s\neq z} } \frac{\sigma_{sz}(v)}{\sigma_{sz}}\kappa(s,z,v)$$
\end{corollary}
$\texttt{p-ab}$ has lower variance than $\texttt{p-rk}$. Intuitively, this is due to the fact that $\texttt{p-ab}$ collects more information per sample compared to the other estimator,  thus its estimations, for the same sample size, are more accurate.
\subsection{Supremum Deviation and Rademacher Averages and non-uniform Bounds}
Here we define the \emph{Supremum Deviation} (SD) and the \emph{c-samples Monte Carlo Empirical Rademacher Average} (c-MCERA). For more details about the topic we refer to the book~\cite{Shalev_2014} and to~\cite{Barlett_2003}. Let $\calD$ be a finite domain and consider a probability distribution $\pi$ over the elements of $\calD$. Let $\calF$ be a family of functions from $\calD$ to $[0,1]$, and $\calS= \{s_1,\dots,s_r\}$ be a collection of $r$ independent and identically distributed samples from $\calD$ sampled according to $\pi$. The SD is defined as 
\begin{align*}
	SD(\calF,\calS) = \sup_{f\in \calF}\left|\frac{1}{r}\sum_{i=1}^r f(s_i)- \CExpec{f}{\pi}\right|
\end{align*}	
The SD is the key concept of the study of empirical processes~\cite{Pollard_2012}. One way to derive probabilistic upper bounds to the SD is to use the \emph{Empirical Rademacher Averages} (ERA)~\cite{Koltchinskii_2001}.  Let $\bm{\lambda}\in \{-1,1\}^{r}$ be a vector or i.i.d. Rademacher random variables, the ERA of $\calF$ on $\calS$ is $$R(\calF,\calS) = \CExpec{\sup_{f\in\calF}\frac{1}{r}\sum_{i = 1}^r \lambda_i f(s_i)}{\bm{\lambda}}$$ Computing the ERA $R(\calF,\calS)$ is usually intractable, since there are $2^r$ possible assignments for $\bm{\lambda}$ and for each such assignment a supremum over $\calF$ must be computed.
In this work we use the state-of-the-art approach to obtain sharp probabilistic bounds on the ERA that uses Monte-Carlo estimation~\cite{Barlett_2003}. Consider a sample $\calS= \{s_1,\dots s_r\}$, for $c\geq 1$ let $\bm{\lambda}\in \{-1,1\}^{c\times r}$ be a $c\times r$ matrix of i.i.d. Rademacher random variables. The c-MCERA of $\calF$ on $\calS$ using $\bm{\lambda}$ is
\begin{align*}
R_r^c(\calF,\calS,\bm{\lambda}) = \frac{1}{c}\sum_{j=1}^{c}\sup_{f\in \calF} \frac{1}{r}\sum_{i = 1}^r \lambda_{j,i}f(s_i)
\end{align*}
The c-MCERA allows to obtain sharp data-dependent probabilistic upper bounds on the SD, as they directly estimate the expected SD of sets of functions by taking into account their correlation. Moreover, they are often significantly more accurate than other methods~\cite{Pellegrina_2022,Pellegrina_2023,Pellegrina_2024}, such as the ones based on loose deterministic upper bounds to ERA~\cite{Riondato_2018}, distribution-free notions of complexity such as the Hoeffding's bound or the VC-Dimension, or other results on the variance~\cite{Maurer_2009,Santoro_2022}. 
Moreover, a key quantity governing the accuracy of the c-MCERA is the \emph{empirical wimpy variance}~\cite{Boucheron_2013} $\mathcal{W}_\calF(\calS)$, that for a sample of size $r$ is defined as 
\begin{align*}
	\mathcal{W}_\calF(\calS) = \sup_{f\in\calF}\frac{1}{r}\sum_{i\in[r]}(f(s_i))^2
\end{align*}

\begin{theorem}
    For $c,r\geq 1$, let $\bm{\lambda}\in \{-1,+1\}^{c\times r}$ be a $c\times r$ matrix of Rademacher random variables, such that $\lambda_{j,i}\in \{-1,+1\}$ independently and with equal probability. Then, with probability at least $1-\delta$ over $\bm{\lambda}$, it holds
    \begin{align*}
        R(\calF,\calS)\leq R_r^c(\calF,\calS,\bm{\lambda}) + \sqrt{\frac{4\mathcal{W}_\calF(\calS)\ln(1/\delta)}{cr}}
    \end{align*}
\end{theorem}
Finally, in this work we use the novel sharp \emph{non-uniform} bound to the SD proposed by Pellegrina and Vadin in~\cite{Pellegrina_2024}. 
\begin{theorem}\label{thm:sup_dev_rade}
	Let $\calF = \bigcup_{i\in [1,t]}\calF_i$ be a family of functions with codomain in $[0,1]$, and let $\calS$ be a sample of $r$ random samples from a distribution $\pi$.  Denote $\hat{v}_{\calF_i}$ such that $\sup_{f\in \calF_i} \Var{f}\leq \hat{v}_{\calF_i}$ for each $i\in [1,t]$. For any $\delta \in (0,1)$, define 
	\begin{align}\label{eq:sd_bound_rade}
		& \tilde{R} = R_r^c(\calF_i,\calS,\bm{\lambda}) + \sqrt{\frac{4 \mathcal{W}_{\calF_i}(\calS)\ln(4t/\delta)}{cr}} \nonumber &\\
		& R_i = \tilde{R_i} + \frac{\ln(4t/\delta)}{r}+\sqrt{\left(\frac{\ln(4t/\delta)}{r}\right)^2 + \frac{2\ln(4t/\delta)\tilde{R_i}}{r}}\nonumber \nonumber& \\
		& \xi_{\calF_i} = 2R_i + \sqrt{\frac{2 \ln(4t/\delta)\left(\hat{v}_{\calF_i}+4R_i\right)}{r}}+\frac{\ln(4t/\delta)}{3r} &
	\end{align}
	With probability at least $1-\delta$ over the choice of $\calS$ and $\bm{\lambda}$, it holds $SD(\calF_i,\calS)\leq \xi_{\calF_i} $ for all $i\in [1,t]$.
\end{theorem}
The idea behind Theorem~\ref{thm:sup_dev_rade} is that when simultaneously bounding the deviation of multiple functions belonging to a set $\calF$, the accuracy of the probabilistic bound on the SD exhibits a strong dependence on the maximum variance $\sup_{f\in\calF }\Var{f}$. Moreover, when the variances of $\calF$'s members are highly heterogeneous, it is better to first partition $\calF$ in $t$ disjoint subsets, where the functions within the same $\calF_i$ share similar variance. This partitioning, leads to sharper bounds on the $SD$ as showed in~\cite{Pellegrina_2024}.
\section{Fast Estimation of the Percolation Centrality}
In this section we provide a new upper bound on the sufficient number of samples to obtain accurate approximations of the percolation centrality. Our result is a generalization of a novel sample-complexity bound for the betweenness centrality~\cite{Pellegrina_2024} that relies on connections between key results from combinatorial optimization and fundamental concentration inequalities.

\subsection{Overall Strategy}
Our strategy is to generalize the results by Pellegrina and Vandin for the betweenness centrality~\cite{Pellegrina_2024} to the percolation centrality. Such approach can be divided in two parts: (1) speeding up the \texttt{p-ab} estimator and (2) derive a better bound on the minimum number of samples needed to achieve a desired \emph{absolute} approximation.
\subsection{Computation of the Percolation Scores}
Here we optimize the graph traversal technique, by generalizing a well known heruistic called \emph{balanced bidirectional BFS} to compute the percolated paths. Furthermore, computing the percolation centrality estimates using \texttt{p-rk} and \texttt{p-ab} works as follows: given two nodes $s$ and $z$, a \emph{truncated} BFS is initialized from $s$ and expanded until $z$ is found. Such approach, produces the set $\Gamma_{sz}$ in time $\bigO(|E|)$. However, we notice that the computation of $\Gamma_{sz}$ can be further ``improved''. Indeed we can generalize the \emph{balanced bidirectional BFS} heuristic by Borassi and Natale in~\cite{Borassi_2019} to speed-up the shortest path sampling procedure to sample percolated shortest paths. Such heuristic speeds-up $\Gamma_{sz}$'s computation to $\bigO(\sqrt{|E|})$ with high probability on several random graphs models, and experimentally on real-world instances. This technique can be applied to \texttt{p-rk}~\cite{Borassi_2019} as well as \texttt{p-ab}~\cite{Pellegrina_2024}. 
More prcisely, given a couple of nodes $s\neq z$, we perform at the same time a BFS from $s$ and a BFS from $z$ until the two BFSs touch each other (in case of a directed graph, we perform a ``forward'' BFS from $s$ and a ``backward'' BFS from $z$). Assume that we visited up to level $\ell_s$ from $s$ and to level $\ell_z$ from $z$, let $N^{(\ell_s)}(s) = \{v \in V : d(s,v)=\ell_{s}\}$ be the set of nodes at distance $\ell_s$ from $s$ and $N^{(\ell_z)}(z)$ be the set of nodes at distance $\ell_z$ from $z$. Then the following simple rule is applied: if $\sum_{u\in N^{(\ell_s)}(s)}deg(u)\leq \sum_{u\in N^{(\ell_z)}(z)}deg(u)$, we perform a new step of the traversal from $s$ by processing all nodes in $N^{(\ell_s)}(s)$, otherwise we perform a new step of traversal from $z$. Intuitively, at each time step, we always process the level that contains less nodes to visit. For the sake of explanation, assume that we are processing $u\in N^{(\ell_s)}(s)$. For each neighbor $w$ of $u$ we do the following:
\begin{itemize}
    \item If $w$ was never visited, add $w$ to $N^{(\ell_s+1)}(s) $;
    \item If $w$ was already visited by $s$, do nothing;
    \item If $w$ was visited by $z$, add the edge $(u,w)$ to the set of candidates $\mathcal{E}$.
\end{itemize}
The parallel traversal stops when at least one between $N^{(\ell_s)}(s) $ and $N^{(\ell_z)}(z)$ is empty (thus, $s$ and $z$ are not connected), or if $\mathcal{E}$ is not empty (thus, $s$ and $z$ met). In the latter case, $\Gamma_{sz}$ between $s$ and $z$ is \emph{implicitly} computed by the two BFSs. As showed in~\cite{Pellegrina_2024}, its very efficient to sample \emph{multiple} shortest paths uniformly at random from $\Gamma_{sz}$. Indeed, it suffices to repeat the following path sampling procedure for an appropriate number of times:
\begin{itemize}
    \item[1.] Sample an edge $(u,v)$ from $\mathcal{E}$ with probability proportional to $\sigma_{su}\sigma_{vz}$;
    \item[2.] Select the path by concatenating a random path from $s$ to $u$, the edge $(u,v)$, and a random path from $v$ to $z$.
\end{itemize}
Moreover, every time a random path $p_{sz}$ is sampled, we add it to a \emph{bag of paths} $\uptau$. To wrap up, the overall sampling procedure can be described as follows:
\begin{description}
	\item[1.] Sample $s$ and $z$ uniformly at random from $V$;
	\item[2.] Run a balanced bidirectional BFS from $s$ and $z$, until the two BFSs meet;
	\item[3.] Sample uniformly at random $\lceil\alpha\sigma_{sz}\rceil$ shortest paths from $\Gamma_{sz}$ where $\alpha\geq 1$ is a positive constant;
\end{description}
Once the bag of shortest paths $\uptau$ is obtained from this sampling procedure, we compute the following function:
\begin{align}\label{eq:estimator}
f_v(\uptau) = \frac{1}{|\uptau|}\sum_{p_{sz}\in \uptau}\mathds{1}\left[v\in p_{sz}\right] \cdot \kappa(s,z,v)
\end{align}

here $\mathds{1}\left[v\in p_{sz}\right] =1 $ if $v\in \texttt{Int}(p_{sz})$ and $0$ otherwise. Moreover, the set of functions we use for the percolation centrality approximation contains all the $f_v$ with $v\in V$ such that $\calF= \{f_v:v\in V\}$. By considering a sample $\calS$ of size $r$ sampled as described above, we define the estimate $\tilde{p}(v)$ of the percolation centrality $p(v)$ of $v$ as
$	\tilde{p}(v) = \frac{1}{r}\sum_{\uptau\in\calS}f_v(\uptau)
$. We have that $\tilde{p}(v)$ is an unbiased estimator of $p(v)$ 
\begin{align*}
&\CExpec{\tilde{p}(v)}{\calS} =\Expec{\frac{1}{|\uptau|} \sum_{p_{sz}\in \uptau}\mathds{1}\left[v\in p_{sz}\right] \cdot \kappa(s,z,v)}  = p(v)&
\end{align*}
As for $\alpha$, using a Poisson approximation to the balls and bins model~\cite{Mitzenmacher_2017} its possible to show that the expected fraction of shortest path that are not sampled from the set $\Gamma_{sz}$ during step $3.$ is $\sigma_{sz}(1-1/\sigma_{sz})^{\alpha\sigma_{sz}} \approx e^{-\alpha}$. Thus, by setting $\alpha = \ln \frac{1}{\beta}$, where $\beta$ is a small value (i.e., $\beta = 0.1$) we obtain that the set of sampled shortest paths is ``close'' to $\Gamma_{sz}$. In other words, this allows to use the balanced bidirectional BFS approach with the \pab~estimator that is preferable to the \prk~(see Corollary~\ref{corollary:variance}).

\subsection{Sample Complexity Analysis}
We propose a generalization of the distribution-dependent bound for the betweenness centrality in~\cite{Pellegrina_2024}, that takes into account the maximum variance of the percolation centrality estimators. Our bound scales with the sum of the percolation centrality (instead of the diameter) of the analyzed graphs. We observe that the sum of the percolation centralities of all nodes in a graph $G$ is upper bounded by the average number of internal nodes $\rho$.
\begin{lemma}\label{lemma:avg_path_length}
	$\sum_{v\in V} p(v) \leq \rho$
\end{lemma}
\begin{proof}
\begin{align*}
&\sum_{v\in V} p(v) = \sum_{v\in V}  \frac{1}{n(n-1)}\sum_{\substack{(s,z)\in V\times V\\ s\neq v\neq z}}{\frac{\sigma_{sz}(v)}{\sigma_{sz}}\cdot \kappa(s,z,v)}  = \frac{1}{n(n-1)}\sum_{v\in V} \sum_{\substack{(s,z)\in V\times V\\ s\neq v\neq z}}{\frac{\sigma_{sz}(v)}{\sigma_{sz}}\cdot \kappa(s,z,v)} &\\& = \frac{1}{n(n-1)} \sum_{\substack{(s,z)\in V\times V\\ s\neq v\neq z}}\frac{1}{\sigma_{sz}}\sum_{p_{sz}\in \Gamma_{sz}}\sum_{v\in V}{\mathds{1}\left[v\in p_{sz}\right] \cdot\kappa(s,z,v)} \leq \frac{1}{n(n-1)} \sum_{\substack{(s,z)\in V\times V\\ s\neq v\neq z}}\frac{1}{\sigma_{sz}}\sum_{p_{sz}\in \Gamma_{sz}}\sum_{v\in V}{\mathds{1}\left[v\in p_{sz}\right] } &
\\& =\frac{1}{n(n-1)}\sum_{s,z\in V} \frac{|\texttt{Int}(p_{sz})|\sigma_{sz}}{\sigma_{sz}} = \frac{1}{n(n-1)}\sum_{s,z\in V}|\texttt{Int}(p_{sz})| = \sum_{v\in V} b(v)&
\end{align*}
where the inequality follows by the fact that
$
\kappa(s,z,v)\leq 1 
$
for all $s,v,z\in V$ such that $s\neq v\neq z$, and $b(v)$ is the betweenness centrality of $v\in V$.
\end{proof}
Moreover, the percolation centrality measure, being built on top of the betweenness, satisfies a form of negative correlation among the vertices of the graph as well: the existence of a node $v$ with high percolation centrality $p(v)$ constraints the sum of the percolation centrality $\Psi = \frac{1}{n(n-1)} \sum_{\substack{(s,z)\in V\times V\\ s\neq v\neq z}}\sum_{v\in V}\frac{\sigma_{sz}(v)}{\sigma_{sz}}{\kappa(s,z,v)} $ of all the nodes to be at most $\Psi - p(v)$; this means that the number of vertices of $G$ with high percolation centrality cannot be arbitrarily large. Unfortunately, we observe that $\Psi$ is an \emph{unknown} parameter since its computation would require the computation of the \emph{exact} percolation centrality for each node $v\in V$. Luckily for us, $\Psi$ is upper bounded by the average number of internal nodes in a shortest path $\rho$, and we can approximate $\rho$ with high precision using well known approximation algorithms (e.g.,~\cite{Boldi_2011,Amati_2023}). In addition, we assume that the maximum variance $\max_{v\in V}\Var{f_v}$ of the percolation centrality estimators is at most a value $\hat{v}$, rather than having a worst-case bound. Thus, the estimates $\tilde{p}(v)$, can exhibit large deviations w.r.t. to the expected value of $p(v)$. Lemma~\ref{lemma:avg_path_length} allows us to generalize the state-of-the-art sample complexity bounds for the betweenness centrality in~\cite{Pellegrina_2024} to the percolation centrality.
\begin{theorem}[Adaptation of Theorem~4.7 in~\cite{Pellegrina_2024}]\label{thm:sample_size}
  Let $\calF = \{f_v : v\in V\}$ be a set of functions from a domain $\calD$ to $[0,1]$. Let a distribution $\gamma$ such that $\CExpec{f_v(\uptau)}{\uptau\sim\gamma} = p(v)$. Define $\hat{v} \in (0,1/4]$, and $\rho\geq 0$
 such that
 \begin{align*}
     & \max_{v\in V} \Var{f_v}\leq \hat{v},\text{ and } \sum_{v\in V}p(v)\leq \Psi
 \end{align*}
    Fix $\delta,\varepsilon\in (0,1)$, and let $\calS$ be an i.i.d. sample of size $r\geq 1$ sampled from $\calD$ according to $\gamma$ such that 
    \begin{align*}
    &r\approx \frac{2\hat{v}+\frac{2}{3}\varepsilon}{\varepsilon^2}\left(\ln\left(\frac{2\Psi}{\hat{v}}\right)+\ln\left(\frac{1}{\delta}\right)\right)\in\bigO\left(\frac{\hat{v}+\varepsilon^2}{\varepsilon^2}\ln\left(\frac{\rho}{\delta\hat{v}}\right)\right)&
     \end{align*}
with probability at least $1-\delta$ over $\calS$, it holds $SD(\calF,\calS)\leq \varepsilon$.
 \end{theorem}
The main difference between the bound in Theorem~\ref{thm:sample_size} and the one for the betweenness centrality in~\cite{Pellegrina_2024} is that it depends on the \emph{maximum variance} of the percolation centrality, that in practice is much smaller than the maximum variance of the betweenness centrality. Finally, we observe that such a bound on the sufficient number of samples needed to obtain an absolute $\varepsilon$ approximation for every node $u\in V$ with probability $1-\delta$, on real world graphs, is much smaller by the one provided by Lima et al.~\cite{Lima_2020}. That is because $\Psi\leq\rho\leq VD(G)$ (see Table~\ref{tab:datasets}).

\subsection{Algorithm Description} In this section we present a randomized algorithm for the approximation of the percolation centrality for every vertex of a given graph. The algorithm is a generalization of the SILVAN algorithm for the betweenness centrality by Pellegrina and Vandin~\cite{Pellegrina_2024}. More precisely, given a graph $G = (V,E)$, the percolation states $x_v$ for each $v\in V$, the quality and confidence parameters $\varepsilon,\delta\in (0,1)$, the number of monte-carlo trials $c$ and the sample size $r'$ used for the bootstrap phase  Algorithm~\ref{algo:apx_algo} works as follows. At the beginning, a bootstrap phase is executed (lines 1-9) in which first the values $\text{minus\_s}[v] = \sum_{u\neq w\neq v}R(x_u-x_w)$ for each vertex $v\in V$ which are necessary to compute $\tilde{p}(v)$, are obtained by running the linear time dynamic programming algorithm proposed in~\cite{Lima_2020}. Next, the algorithm runs $r' = \ln\left(1/\delta\right)/\varepsilon$ bidirectional balanced BFSs and generates a set of shortest paths $\calS'$ that is then used to partition $\calF$ in $t$ subsets. Such partition is obtained by using the efficient \emph{empirical peeling scheme} introduced in~\cite{Pellegrina_2024} (lines 3-4). The idea of the latter approach is to partition $\calF$ into subsets of functions that share similar variance. Such approach allows to control the supremum deviations $SD(\calF_j,\calS)$ for each $\calF_j$ separately exploiting the fact that the maximum variance is computed on each subset $\calF_j$ instead of the whole set $\calF = \bigcup_{j=1}^t\calF_j$. This leads to sharp non-uniform bounds that are locally valid for each subset $\calF_j\subseteq \calF$. Moreover, we refer to~\cite{Pellegrina_2024} (Section 4.2) for a more detailed description of the method, for our purpose we can use their approach as a sort of ``black-box'' tool to improve the bounds on the c-MCERA. In addition, while drawing random shortest paths from $G$ (line 3), the algorithm, computes an approximation of the average internal path length $\rho$. Next, Algorithm~\ref{algo:apx_algo}, computes the upper bound on the number of samples $r$ according to Theorem~\ref{thm:sample_size}, i.e. sets the overall sample size to $r= \bigO\left(\frac{\tilde{\hat{v}} +\varepsilon^2}{\varepsilon^2}\ln\left(\frac{\tilde{\rho}}{\delta \tilde{\hat{v}}}\right)\right)$ where $\tilde{\hat{v}}$ is the estimated maximum variance over all the $\calF_j$ and $\tilde{\rho}$ is an estimate of $\rho$ computed in lines 3-4. Moreover, as a final step of the bootstrap phase, the \emph{geometric sampling schedule} is defined. The schedule is chosen such that  the sample size $r_i$ is increased according to a geometric progression: $r_{i+1} = r_i \cdot x$ where $x=1.2$. Finally, the set of confidence parameters $\{\delta_i = \delta/2^{i+1}\}$ is chosen. Such confidence parameters for the scheduling is chosen such that $\sum_{i\geq 1}\delta_i \leq \delta/2$. Moreover, after the choice of the sample schedule, the approximation phase starts. As a first step of the second phase, all the $\xi_{\calF_j}$ are set to $1$, and other variables are initialized (line 7-9). In every iteration of the while loop (lines 10-13), the algorithm executes the following operations: it increments the iteration index $i$, adds $d_r = r_i-r_{i-1}$ new samples to $\calS_i$ using \texttt{sampleSPs}($d_r$), then it adds $d_r$ new columns of length $c$ to the matrix $\bm{\lambda}$ so that $\bm{\lambda}\in\{0,1\}^{c\times r_i}$. These new columns are generated by sampling a $c\times d_r$ matrix in which each entry is a Rademacher random variable. The algorithm updates all the estimates using the procedure \texttt{up.eEstimates}. Such procedure, uses the sample $\calS_i$ the matrix $\bm{\lambda}$, the partition $\{\calF_j\}$, and the array $\{\text{minus\_s}[v], \forall v \in V \}$ to compute: a vector of percolation centrality estimates $\tilde{p}(v)$ for each $v\in V$ a $n\times c$ matrix $\tilde{R}$ in which each entry $\tilde{R}(v,k)$ is the estimated c-MCERA for the function $f_v$ using $\bm{\lambda}$'s $k^{th}$ row $\bm{\lambda}_{k,\cdot}$ such that $\tilde{R}(v,k) = R^c_{r_i}(\{f_v\},\calS_i,\lambda_{k,\cdot})$, these values are needed to compute the c-MCERA of each set $\calF_i$. Then the set $\{\hat{v}_{\calF_i}\}$ contains probabilistic upper bounds to the supremum variances. The procedure \texttt{up.eEstimates} increases the value of $\tilde{p}(v)$ for all $v\in p_{sz}$ and for all $p_{sz}\in \uptau$ according to the estimator described in Equation~\ref{eq:estimator}. Moreover, in a similar way, it computes $\tilde{R}(v,k)$ for all $k\in [1,c]$ and $\mathcal{W}_{\calF_j}(\calS_i)$ for all $j$ as each new sample is obtained. 
After $\calS_i$ is realized and processed, the algorithm computes (line 14), for all partitions $\calF_j\subseteq\calF$, the c-MCERA $R^c_{r_i}(\calF_j,\calS_i,\bm{\lambda})$ using the values stored in $\tilde{R}$. Subsequently, it computes (line 15) an upper bound $\xi_{\calF_j}$ to the supremum deviation $SD(\calF_j,\calS)$ for each partition $\calF_j$ using the function \texttt{epsBound}. In practice we compute Equation~\ref{eq:sd_bound_rade} by replacing $4/\delta $ with $5/\delta_i$, and by using the quantities computed so far. Finally, \texttt{stoppingCond.} is invoked. In such routine we check whether $\varepsilon\geq \xi_{\calF_j}$ or all $j\in [1,t]$ or if $r_i\geq r$. When one of the two criteria is met, the algorithm exits the while loop and returns the approximation $\tilde{p}$ (line 16).

\begin{theorem}
	Algorithm~\ref{algo:apx_algo} computes an absolute $\varepsilon$ approximation of the percolation centrality of each node with probability at least $1-\delta$ in time $\bigO\left((r+r') (n+m)\right)$ time and using $\bigO(n+m)$ space.
\end{theorem}

\begin{proof}
	The algorithm terminates when \texttt{stoppingCond.} is \emph{true}. This happens when at least one of the two stopping conditions is met, i.e., (1) the algorithm sampled $r_i\geq r$ samples, or (2) the upper bounds on the supremum deviations $\xi_{\calF_j}$ are at most $\varepsilon$ for all $j\in [1,t]$. In both cases, we are guaranteed to have $|p(v)-\tilde{p}(v)|\leq \varepsilon$ for all $v\in V$ with probability of at least $1-\delta$. Finally, the running time of the algorithm is determined by the overall number of bidirectional BFS i.e.,  the running time of the algorithm is the sum between the bootstrap and estimation phases running times. Thus it is $\bigO(r'(n+m))+(r(n+m)) = \bigO\left((r+r') (n+m)\right)$.
\end{proof}

\begin{algorithm2e}[htb!]
\caption{PecolationCentralityApproximation}\label{algo:apx_algo}
\KwIn{Graph $G = (V,E)$, percolation states array $\bm{x}$ s.t. $0\leq x_v\leq 1$ $\forall v\in V$, Monte-Carlo Trials $c\geq 1$, bootstrap iterations $r'\geq 1$,  $\varepsilon,\delta\in (0,1)$. }
\KwOut{$\varepsilon$-apx. of $p(v)$, $\forall v\in V$ w.p. $1-\delta$}

$\texttt{SortAscending}(\bm{x})$\tcp{Sort perc. states}

$\textit{minus}_{\_s} = \texttt{PercolationDifferences}(\bm{x},|V|)$

$\calS' = \texttt{SampleSPs}(r')$

$\{\calF_i, i \in [1,t]\} = \texttt{Emp.Peeling}(\calF,\calS',\textit{minus}_{\_s})$

$r = \texttt{SufficientSampleSize}(\calF,\calS',\delta/2)$

$\{r_i\},\{\delta_i\} = \texttt{samplingSchedule}(\calF,\calS')$

\For{$j\in [1,t] $}{ $\xi_{\calF_j} = 1$}

$i = 0; \calS_0 = \emptyset; \bm{\lambda} = \text{ empty matrix};$

\While{$not\texttt{ stoppingCond.}(\varepsilon,\{\xi_{\calF_j}\},r,r_i)$}{
$i = i+1; d_r =r_i - r_{i-1};$

$\calS_i = \calS_{i-1}\cup\texttt{sampleSPs}(d_r)$
$\bm{\lambda}=\texttt{add columns}(\texttt{sampleRrvs}(d_r,c))\text{ to } \bm{\lambda}$
$\tilde{p},\tilde{R},\{\hat{v}_{\calF_{j}}\} = \texttt{up.eEstimates}(\calS_i,\bm{\lambda},\{\calF_j\},\textit{minus}_{\_s})$

\For{$j\in [1,t]$}{
    $\tilde{R}^c_{r_i}(\calF_j,\calS_i,\bm{\lambda})=\frac{1}{c}\sum_{k = 1}^c\max_{v\in V}\{\tilde{R}(v,k)\}$

    $\xi_{\calF_j} = \texttt{epsBound}(\tilde{R}^c_{r_i}(\calF_j,\calS_i,\bm{\lambda}),\hat{v}_{\calF_j},\delta_i)$

}
}
\Return $\{(v,\tilde{p}(v), v\in V\}$

\end{algorithm2e}

\section{Experimental Evaluation}\label{sec:experimental}
In this section, we summarize the results of our experimental study on approximating
the percolation centrality in real-world networks. We compare our algorithm with the state-of-the-art algorithms to approximate such a centrality measure. 

\begin{table}[htb!]
\setlength{\tabcolsep}{0.2pt} 
\renewcommand{\arraystretch}{1.5} 
\centering
\caption{The data sets used in our evaluation, where $n$ denotes the number of nodes, $m$ the number of edges, $D$ the exact diameter, $\rho$ the exact average internal path length, and $\Psi$ the sum of the exact percolation centralities (type D stands for directed and U for undirected). Dashed lines indicate that the exact values are not available due to the dimension of the data set. In such case, we exhibit an approximation of the $D$ and $\rho$. The approximation is computed by running ten times the sampling algorithm in~\cite{Amati_2023} with $1024$ seed nodes. For such measures, we provide the average estimated value and its standard deviation.}\label{tab:datasets}
\begin{tabular}{lcccccc}
\bottomrule
\multicolumn{1}{l}{\textbf{Graph}} & \textbf{n} & \textbf{m} & \textbf{D} & $\bm{\rho}$ & $\bm{\Psi}$ & \textbf{Type} \\ \hline
\texttt{Gnutella 31}                         & 62586          & 147892         & 31                & 8.20         & 2.96E-05              & D             \\
\texttt{Musae Fb.}                          & 22470          & 170823         & 15                & 3.97         & 0.0002                & U             \\
\texttt{Enron}                              & 36692          & 183831         & 13                & 3.03         & 6.97-05               & U             \\
\texttt{CA-AstroPH}                         & 18771          & 198050         & 14                & 3.19         & 0.0002                & U             \\
\texttt{Cit-HepPh}                         & 34546          & 421534         & 49                & 10.69        & 0.0003                & D             \\
\texttt{Epinions}                           & 75879          & 508837         & 16                & 3.76         & 2.31E-05              & D             \\
\texttt{Slashdot}                           & 82168          & 870161         & 13                & 3.14         & 3.31E-05              & D             \\
\texttt{Web-Not.dame}                     & 325729         & 1469679        & 93                & 10.27        & 5.30E-06              & D             \\
\texttt{Web-Google}                        & 875713         & 5105039        & 51                & 10.71        & 5.89E-06              & D             \\
\texttt{Twitch-Edges}                       & 168114         & 6797557        & 8                 & 1.88         & 1.12E-05              & U \\     \cline{1-7}
\texttt{Wiki Talk} & 2394385&5021410 & 8.2 ($\pm 0.42$) & 2.99 ($\pm 0.05$) & --& D\\ 
\texttt{Twitter} &2541739 & 13708316 &49 ($\pm 1.1$) &5.45 ($\pm 0.08$) & --& D\\ 
\texttt{Flickr} &1715254 &15551249 & 20 ($\pm 0.9$) &4.28 ($\pm 0.04$) & --& U\\ 
\texttt{soc-Pokec} &1632803 & 30622564&15 ($\pm 0.8$) &4.24 ($\pm 0.02$) &-- & D

\\ \bottomrule 
\end{tabular}
\end{table}
\subsection{Networks}
We evaluate all algorithms on real-world graphs of different
nature, whose properties are summarized in Table~\ref{tab:datasets}. These networks come from several domains available on the well known SNAP~\cite{snapnets} repository. We executed the experiments on a server running Ubuntu 16.04.5 LTS equipped with
AMD Opteron 6376 CPU (2.3GHz) for overall 32 cores and 64 GB of RAM. 
\subsection{Implementation and Evaluation details}
We implemented all the algorithms in \texttt{Julia} exploiting parallel computing~\footnote{GitHub code: \url{https://anonymous.4open.science/r/percolation_centrality-892B/}}. For the sake of fairness, we re-implemented the exact algorithm~\cite{Piraveenan_2013} and the approximation ones proposed by Lima et al.~\cite{Lima_2020,Lima_2022}, allowing for them to scale well on the tested graphs (that have higher magnitude compared to the one used in~\cite{Lima_2020,Lima_2022}). Indeed, the exact~\cite{Piraveenan_2013} and the \texttt{p-rk} algorithms original implementations are in \texttt{Python} without exploiting parallelism. While, the \texttt{p-ab} algorithm is not available for download. For every graph, we ran all the algorithms with parameters $\varepsilon\in\{0.1,0.07,0.05,0.01,0.005\}$ and $\delta = 0.1$. Moreover, for the c-MCERA we use $c = 25$ Monte Carlo trials as suggested in~\cite{Cousins_2023,Pellegrina_2024}.  In all the experiments, we set the percolation state $x_v$ fir each $v\in V$, as a random number between $0$ and $1$. Finally, each experiment has
been ran 10 times and the results have been averaged.

\subsection{Experimental Results}
\paragraph{Running times and sample sizes.}

\begin{figure}[htb!]
	\captionsetup[subfigure]{justification=centering}
	\begin{subfigure}{0.45\textwidth}
		\centering
		\includegraphics[scale=0.3]{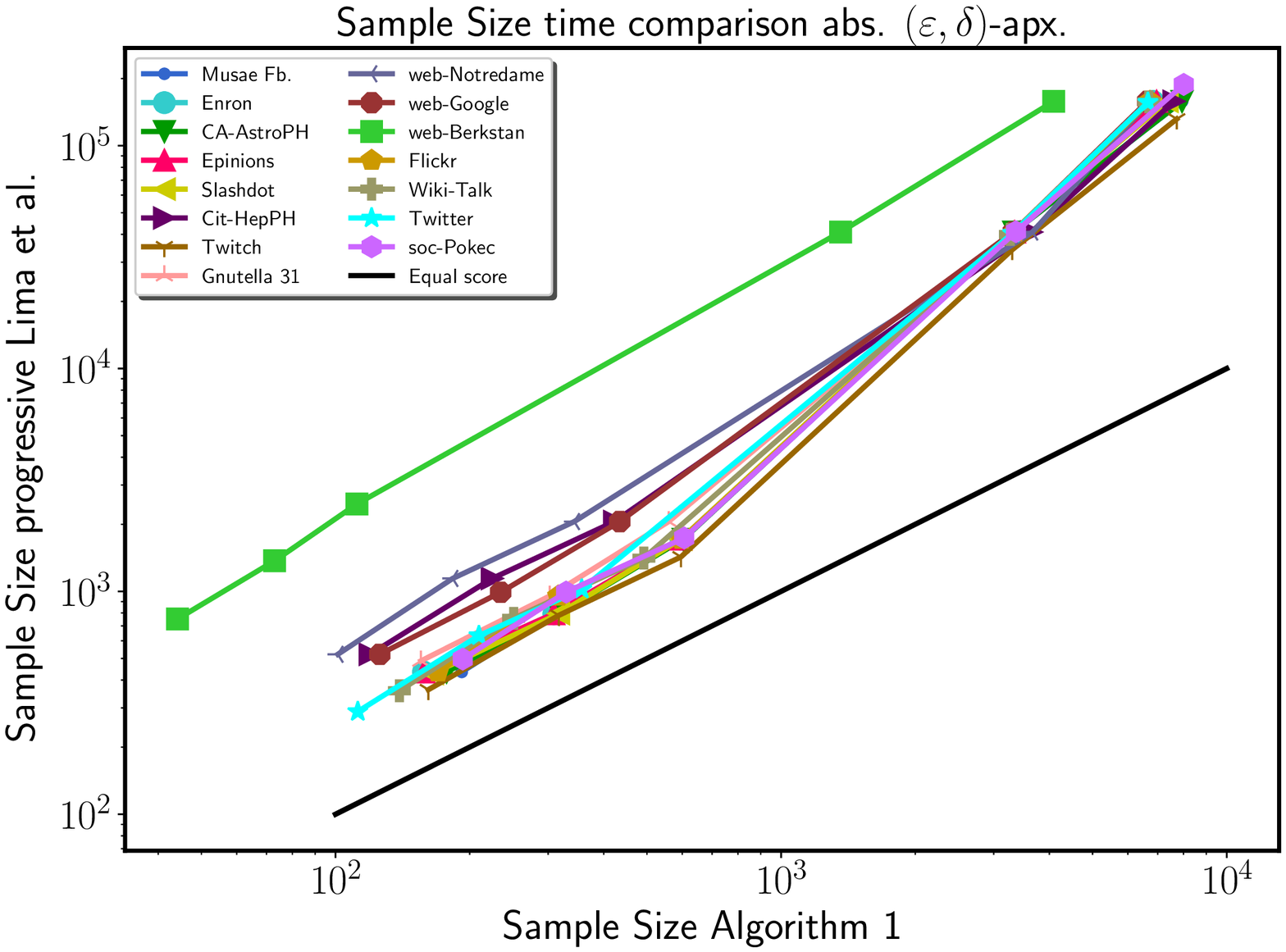}
		\caption{}\label{fig:era_ss}
	\end{subfigure}
	\begin{subfigure}{0.45\textwidth}
		\centering
		\includegraphics[scale=0.3]{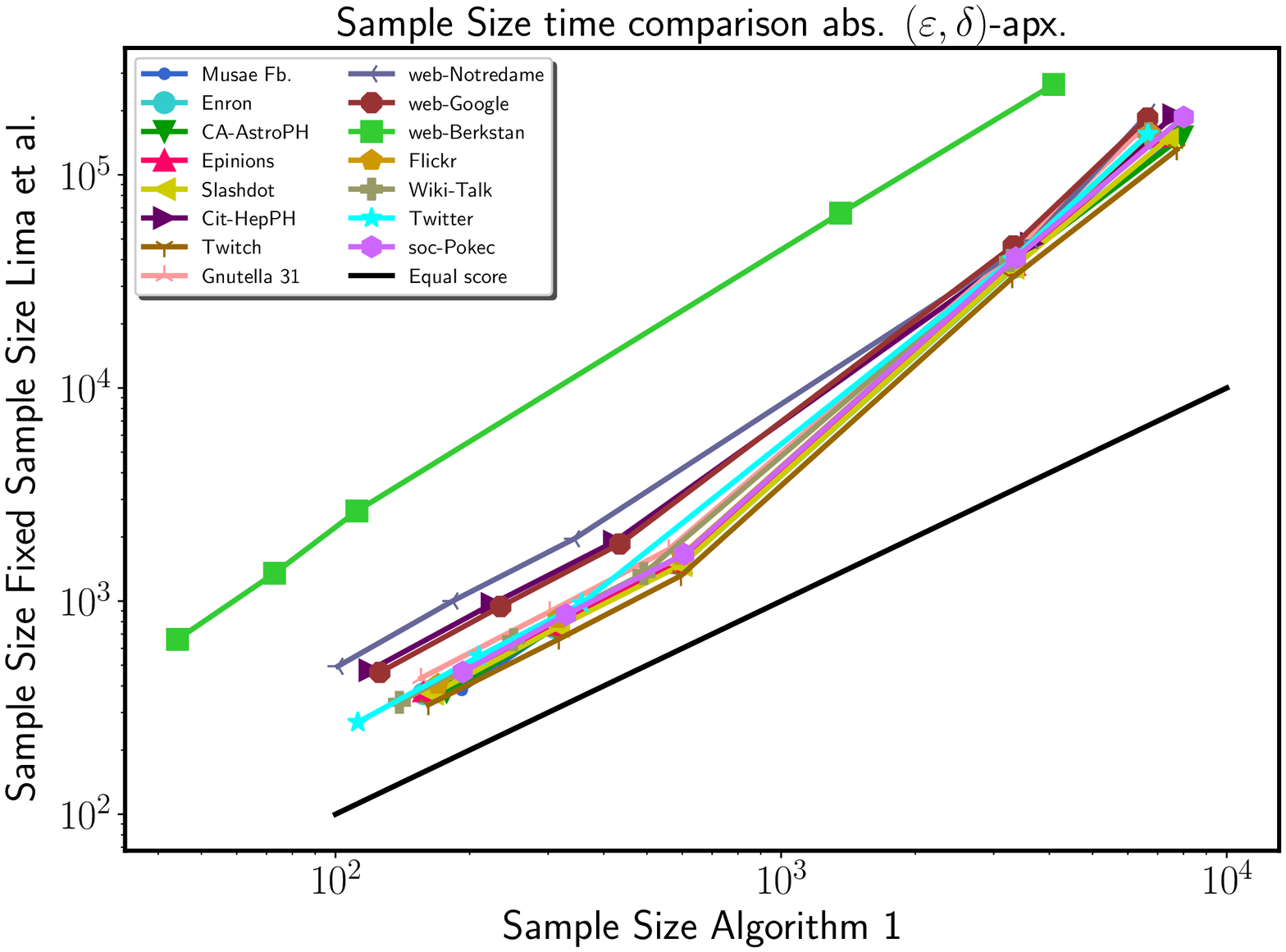}
		\caption{}\label{fig:fixed_ss_s}
	\end{subfigure}
	\caption{Sample size comparison between Algorithm~\ref{algo:apx_algo} and: (a) the progressive sampling algorithm by Lima et al.,~\cite{Lima_2022} and (b) the fixed sample size algorithm by Lima et al.~\cite{Lima_2020}.}\label{fig:sample_size}
\end{figure}
\begin{figure}[htb!]
	\captionsetup[subfigure]{justification=centering}
	\begin{subfigure}{0.45\textwidth}
		\centering
		\includegraphics[scale=0.3]{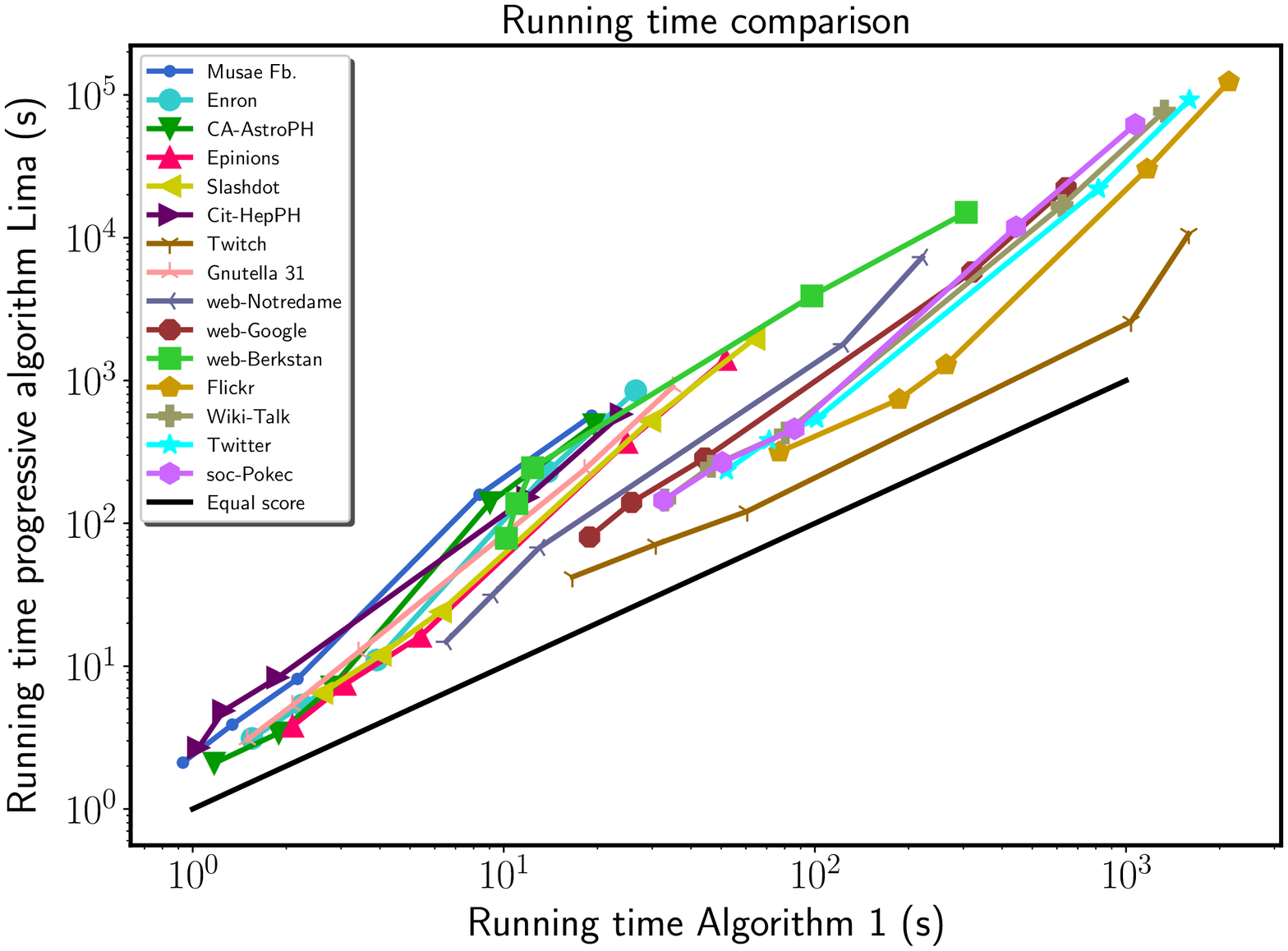}
		\caption{}\label{fig:era_boot}
	\end{subfigure}
	\begin{subfigure}{0.45\textwidth}
		\centering
		\includegraphics[scale=0.3]{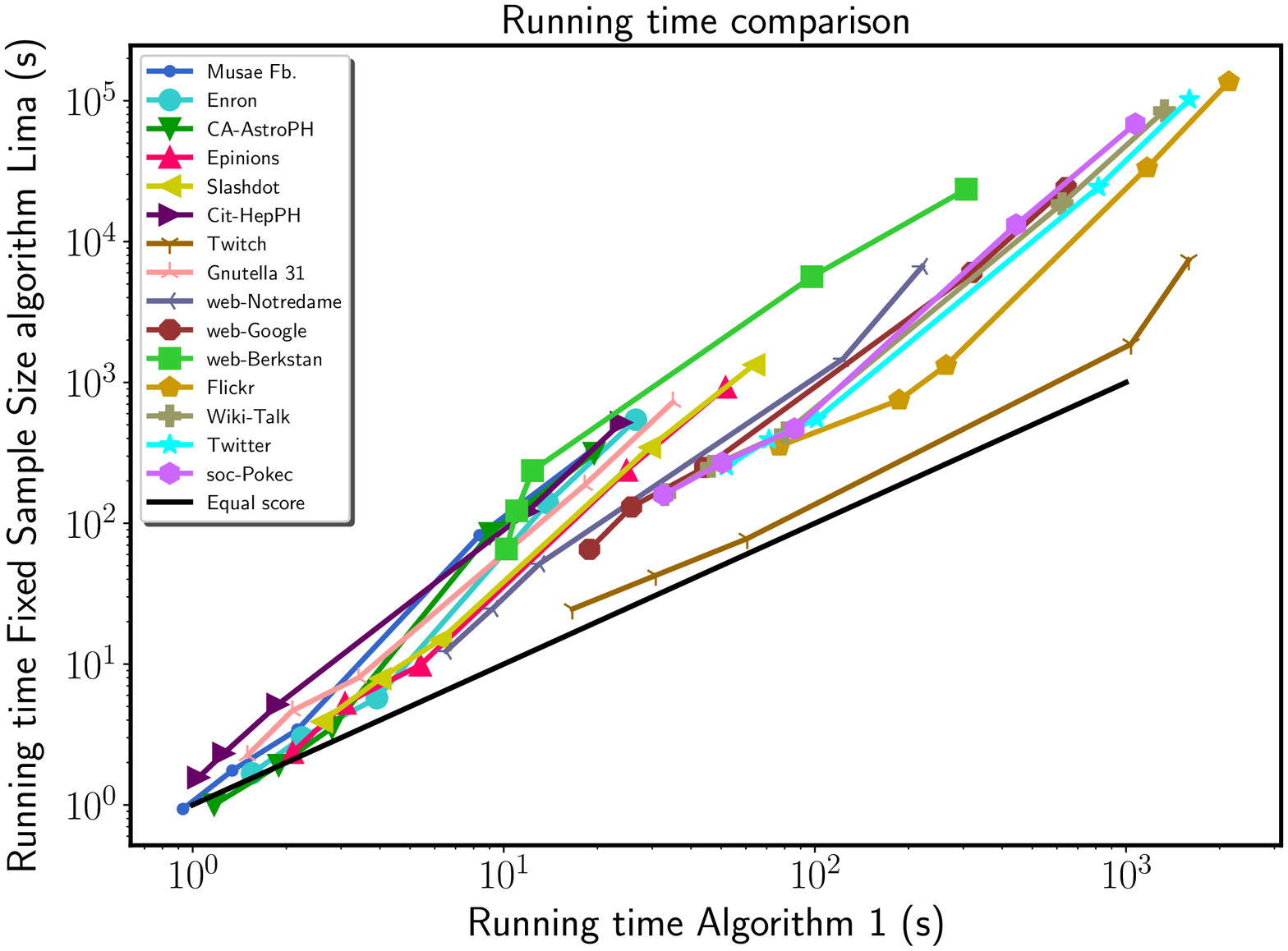}
		\caption{}\label{fig:fixed_ss_boot}
	\end{subfigure}

	\caption{Experimental analysis for $\varepsilon\in \{0.1,0.07,0.05,0.01,0.005\}$. Comparison between \textbf{(a)} the running times of Algorithm~\ref{algo:apx_algo} and the progressive sampling algorithm by Lima et al.~\cite{Lima_2022}; \textbf{(b)} Algorithm~\ref{algo:apx_algo} and the fixed sample size approach by Lima et al.~\cite{Lima_2020};  In figures \textbf{(a-b)} the value of $\varepsilon$ are sorted in descending order.}\label{fig:running_time}
\end{figure}

\begin{figure}[htb!]
	\centering
	\includegraphics[scale=0.3]{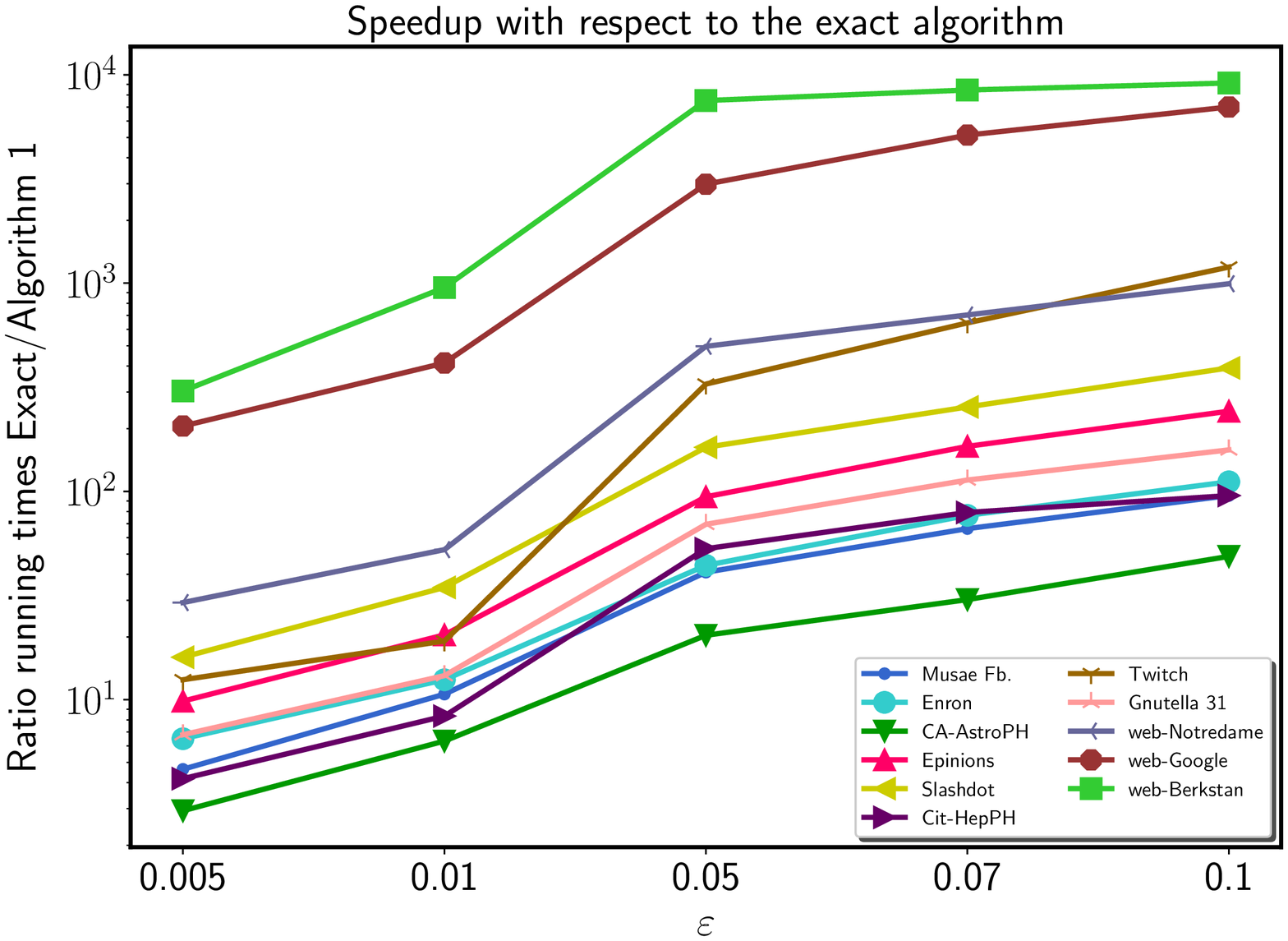}
	\caption{Running time of Algorithm~\ref{algo:apx_algo} and the exact algorithm in terms of speedup to compute the percolation centrality scores. On the $y$ axes we show the ratio between the running time of the exact algorithms and Algorithm~\ref{algo:apx_algo} for different values of $\varepsilon$.}\label{fig:exact}
\end{figure}

In our first experiment, we compare the average sample sizes and execution times of Algorithm~\ref{algo:apx_algo} with the state-of-the-art ones~\cite{Lima_2020,Lima_2022}.

Figure~\ref{fig:sample_size} shows the comparison between the sample sizes needed by Algorithm~\ref{algo:apx_algo} and the state-of-the-art algorithms to approximate the percolation centrality. More precisely, Figure~\ref{fig:era_ss} shows the sample size needed by Algorithm~\ref{algo:apx_algo} ($x$ axes) and the progressive one in~\cite{Lima_2022} ($y$ axes). Moreover, Figure~\ref{fig:fixed_ss_s} shows the same type of comparison between our method and the fixed sample size algorithm in~\cite{Lima_2020}. We observe that, in both cases, Algorithm~\ref{algo:apx_algo} uses fewer samples to compute the absolute $\varepsilon$ approximation of the percolation centrality of each vertex. Indeed, the sample size needed by the other algorithms quickly increases as $\varepsilon$ decreases, up to the point of being $10^5$. This discrepancy in the number of samples needed to converge between our algorithm and the state-of-the-art ones impacts the running times of the considered algorithms. Indeed, our novel algorithm constantly outperforms the progressive sampling one based on loose deterministic upper bounds on the ERA in~\cite{Lima_2022} (see Figure~\ref{fig:era_boot}) and the fixed sample size algorithm~\cite{Lima_2020} (see Figure~\ref{fig:fixed_ss_boot}, and Table~$2$ in the Supplementary Material). In Figure~\ref{fig:running_time} we can see that both the methods by Lima et al. quickly become impractical as the value of $\varepsilon$ decreases. Interestingly, on small graphs for big values of $\varepsilon$ (i.e., $0.1$ and $0.07$), the running times of the fixed sample size algorithm and our approach are very close. This phenomenon is due to the fact that the sample sizes do not differ significantly (see Figure~\ref{fig:fixed_ss_s}) and that in Algorithm~\ref{algo:apx_algo} we ``pay'' some additional time for the bootstrap phase and at each iteration of the while loop (lines 10-15) to check whether the stopping condition is met and to update the estimates on the average number of internal nodes $\tilde{\rho}$. While for the fixed sample size approach we do not have to optimize any function to understand if the convergence criterion is met. Indeed, such method executes $r$ parallel truncated BFS visits, and converges when all the $r$ traversal are completed.

Figure~\ref{fig:exact} shows the ratio between the running times of our approach and the exact algorithm to compute the percolation centrality. For small values of $\varepsilon$ (i.e., $\varepsilon = 0.005,0.01$) Algorithm~\ref{algo:apx_algo} is at least $10$x and $10^2$x faster than the exact algorithm respectively on small and big graphs (such as \texttt{Web-Google} and \texttt{Web-Berkstan}). Moreover, we observe that for bigger values of $\varepsilon$, Algorithm~\ref{algo:apx_algo} has a speedup of at least $10$x and $10^4$x for the analyzed graphs. This result along with the one presented in the next paragraph, suggest that Algorithm~\ref{algo:apx_algo} can be used to obtain very precise estimates of the percolation centrality of each node $u\in V$ on big graphs for which the exact algorithm (or the other methods~\cite{Lima_2020,Lima_2022}) would need an unreasonable amount of time.

\begin{figure}[htb!]
	\centering
	\includegraphics[scale=0.35]{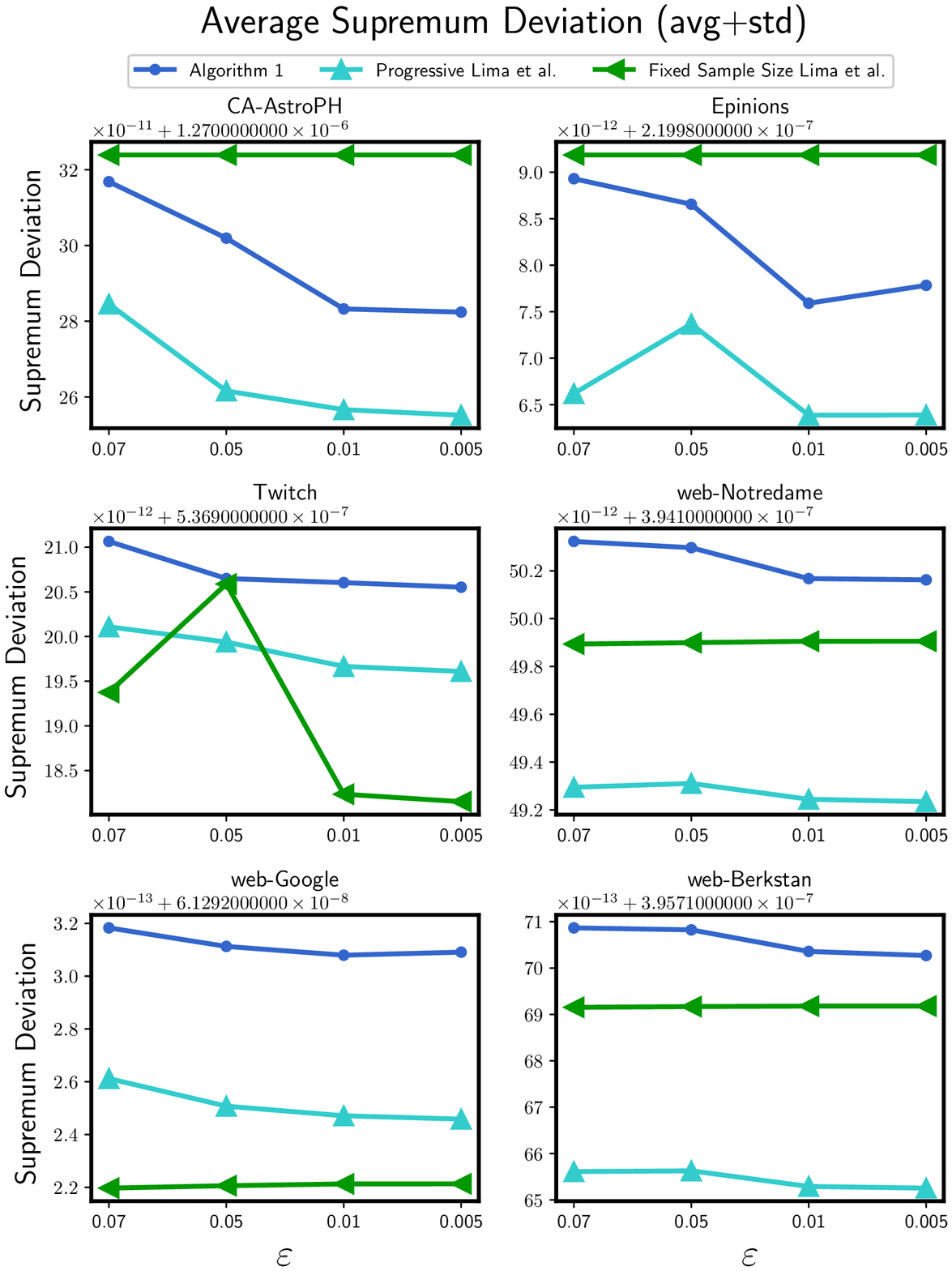}
	\includegraphics[scale=0.35]{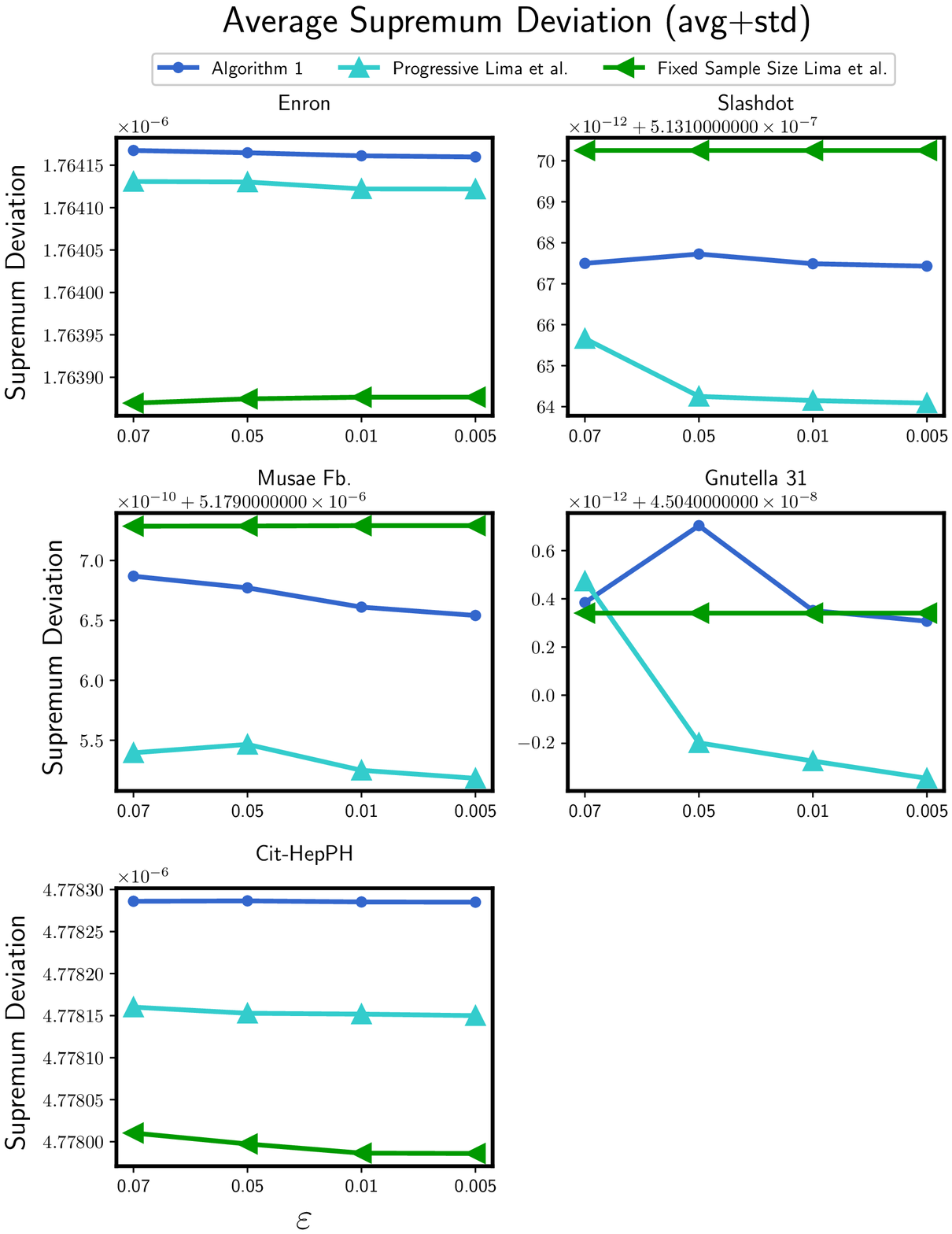}
	\caption{Average Supremum Deviations for the algorithms by Lima et al.~\cite{Lima_2020,Lima_2022} and Algorithm~\ref{algo:apx_algo}. All the plots show the average SDs plus their standard deviation for $\varepsilon\in\{0.07,0.05,0.01,0.005\}$}\label{fig:errors}
\end{figure}

\paragraph{Errors.}
As a second experiment, we compare the quality of the estimates in terms of average supremum deviation of the analyzed algorithms. Moreover, Figure~\ref{fig:errors} shows the average supremum deviation of the percolation centrality estimates of the algorithms by Lima et al.~\cite{Lima_2020,Lima_2022} and Algorithm~\ref{algo:apx_algo} for a subset of the graphs in Table~\ref{tab:datasets} and values of $\varepsilon\in \{0.07,0.05,0.01,0.005\}$. We observe that all the algorithms provide SDs below the desired $\varepsilon$. Surprisingly, all the supremum deviations are at most $10^{-11}$ even for $\varepsilon = 0.07$. This suggests that all the tested algorithms are well suited to compute an absolute $\varepsilon$ approximations for the percolation centrality of all the nodes. Moreover, the algorithms by Lima et al.,~\cite{Lima_2020,Lima_2022} exhibit the smallest SDs on almost all the graphs. This is not surprising, given that, in general, they use bigger sample sizes than Algorithm~\ref{algo:apx_algo} (see Figure~\ref{fig:sample_size}). Such a result suggests that a sample size derived by using the bound in Theorem~\ref{thm:sample_size} is enough to compute sharp estimates of the percolation centrality for each node in a given graph. 

\section{Conclusion}
We presented a randomized algorithm for estimating the percolation centrality of all nodes in a graph. The algorithm relies on the state-of-the-art bounds on the supremum deviation of functions based on the c-MCERA to provide a probabilistically guaranteed absolute $\varepsilon$ approximation of the percolation centrality. Our experimental results (summarized in Section~\ref{sec:experimental}) show the performances of our algorithm versus the state-of-the-art algorithms~\cite{Lima_2020,Lima_2022}. Our approach consistently over-performs its competitors in terms of running time and sample size. Moreover, on all the tested real world graphs, the estimation error in our experiments is many orders of magnitude smaller than the theoretical worst case guarantee. For future work, would be interesting to extend the concept of percolation centrality to group of nodes~\cite{Pellegrina_2023}, and to extend our results to temporal~\cite{Buss_2020} and uncertain graphs~\cite{Saha_2021}.

\bibliographystyle{ACM-Reference-Format}

\bibliography{bibliography}

\appendix
\begin{center}
	\LARGE{\textbf{Appendix}}
\end{center}

\section{Missing proofs}


\paragraph{\textbf{Proof of Theorem~\ref{thm:sample_size}.}}
The proof of Theorem~\ref{thm:sample_size} is a straightforward adaptation of the one for the betweenness centrality provided by Pellegrina and Vandin in~\cite{Pellegrina_2024}. For completeness, we provide the adapted proof.
\begin{proof}
     Define the functions $g(x) = x(1-x)$ and $h(x) = (1+x)\ln (1+x)-x$ for $x\geq 0$. Moreover, ler $\hat{x}_1,\hat{x}_2$, and $\hat{x}$ be
     \begin{align*}
         &\hat{x}_1= &\\&\inf\left\{ x:\frac{1}{2}-\sqrt{\frac{\varepsilon}{3}-\frac{\varepsilon^2}{9}}\leq x\leq \frac{1}{2},g(x)h\left(\frac{\varepsilon}{g(x)}\right)\leq 2\varepsilon^2\right\}&\\&
         \hat{x}_2 = \frac{1}{2}-\sqrt{\frac{1}{4}-\hat{v}},\;\hat{x} = \min\{\hat{x}_1,\hat{x}_2\}&
     \end{align*}
    For a sample $\calS$ of size $r$, define the events $E$ and $E_v$ as 
    \begin{align*}
        &E = \text{`` }\exists v\in V : \left| p(v)-\tilde{p}(v) \right|> \varepsilon\text{ ''}&\\&
        E_v = \text{`` }\left| p(v)-\tilde{p}(v) \right|> \varepsilon\text{ ''}&
    \end{align*}
    Using the union bound, we obtain
    \begin{align*}
        \Prob{E} = \Prob{\bigcup_{v\in V}E_v}\leq \sum_{v\in V} \Prob{E_v}
    \end{align*}
    Next, by applying the Hoeffding's and Bennet's inequalities (cita), Bathia and Davis inequality on variance (cita), and from the fact that $\sup_{f_v\in\calF}\Var{f_v}\leq \hat{v}$, it holds, define 
    
    \begin{align*}
	&A(v,r,\varepsilon) = &\\& \exp\left(-r\min\{\hat{v},g(p(v))h\left(\frac{\varepsilon}{\min\{\hat{v},g(p(v))\}}\right)\}\right)&
    \end{align*}
	 for all $v\in V$
    \begin{align*}
    \Prob{E_v}\leq 2\min\left\{\exp\left(-2r\varepsilon^2\right),A(v,r,\varepsilon) \right\}    
    \end{align*}
    Define the functions $H(r,\varepsilon) = \exp\left(-2r\varepsilon^2\right)$, $B(x,r,\varepsilon) = \exp(-rxh(\varepsilon/x)))$ and $\phi(x)$ (defined below), we write
    \begin{align}
        &\Prob{E}\leq \sum_{v\in V} 2 \min\{H(r,\varepsilon),B(\min\{\hat{v},g(p(v))\},r,\varepsilon\} &\\ &=  \sum_{v\in V}\phi(p(v))\label{eq:prob_1}&
    \end{align}
    We observe that the values of $p(v)$ are unknown a priori, and that it is not possible to directly compute the r.h.s. of the equation above. We can obtain a sharp upper bound by leveraging constraints on the possible values of $p(v)$ imposed by $\hat{v}$ and $\Psi$. To do so, we define an appropriate optimization problem with respect to the values of $p(v)$. Let $r_x$ be the number of nodes of $V$ that we assume have $p(v)=x$, for $x\in (0,1)$ (nodes $v$ such that $p(v) = 0 $ or $p(v) = 1$ can be safely ignored since $f_v$ is constant and $\Prob{E_v} = 0$); then, we formulate the following constrained optimization problem over the variables $r_x$:
    \begin{align*}
\begin{array}{ll@{}ll}
\text{maximize}  & \displaystyle\sum\limits_{x\in (0,1)} r_x \cdot \phi(x) &\\
\text{subject to}& \displaystyle\sum\limits_{x\in (0,1)}   x\cdot r_x \leq \Psi,  &\\
                 &0\leq r_x\leq \frac{\Psi}{x} ,r_x\in \mathbb{N}&
\end{array}
\end{align*}
The first constrain follows from $\sum_{v\in V}p(v)\leq \Psi$, while the second set of constraints imposes that $r_x$ are positive integers and that there cannot be more than $\Psi/x$ nodes with $p(v) = x$ by definition of $\Psi$. Thus, from Equation~\ref{eq:prob_1}, the value of the objective function of the optimal solution o this problem upper bounds $\Prob{E}$, as we consider the worst-case configuration of the admissible values of $p(v)$. Moreover, this formulation is a specific instance of the Bounded Knapsack Problem~\cite{Martello_1990} over the variables $r_x$, where items with label $x$ are selected $r_x$ times, with unitary profit $\phi(x)$ and weight $x$. Furthermore, each element can be selected at most $\Psi/x$ times, while the total knapsack capacity is $\Psi$. As in~\cite{Pellegrina_2024}, we are not interested in the optimal solution of the integer problem, rather in its upper bound given by the optimal solution of the continuous relaxation in which we let $r_x\in\mathbb{R}$. Informally, the solution is obtained by choosing at maximum capacity every item in decreasing order of profit-weight ratio $\phi(x)/x$ until the total capacity is saturated. In our case, it is enough to fully select the item with higher profit-weight ratio to fill the entire knapsack. Formally, define 
\begin{align*}
    x^\star = \argmax_{x\in (0,1)}\left\{\frac{\phi(x)}{x}\right\}
\end{align*}
the optimal solution to the continuous relaxation is $r_{x^\star} = \Psi/x^\star$, $r_x = 0$ for all $x\neq x^\star$, while the optimal objective is equal to 
\begin{align*}
    \frac{\Psi\phi(x^\star)}{x^\star}\geq \Prob{E}
\end{align*}
Observe that $x^\star$ always exists, as $\phi(x)/x$ is a positive function in $(0,1)$. The search of $x^\star$ can be simplified by exploiting the same approach used in~\cite{Pellegrina_2021} for the betweenness centrality, and leads to 
\begin{align*}
	&\Prob{E}\leq \sup_{x\in (0,\hat{x})}\left\{\frac{\Psi 2B(g(x),r,\varepsilon)}{x}\right\} &
\end{align*}
Setting $r\geq \sup_{(0,\hat{x})}\left\{\ln(\frac{2\Psi^{(\star)}}{x\delta})/(g(x)h(\frac{\varepsilon}{g(x)}))\right\}$ it holds that $\Prob{E}\leq \delta$. In order to approximate $r$, the r.h.s. of the equation can be computed using a numerical procedure~\cite{Pellegrina_2024} obtaining the following approximation: 
\begin{align*}
    r\approx\frac{2\hat{v}+\frac{2}{3}\varepsilon}{\varepsilon^2}\left(\ln\left(\frac{2\Psi}{\hat{v}}\right)+\ln\left(\frac{1}{\delta}\right)\right)
\end{align*}
We observe that the sample size $r$ depends on the sum of the percolation centralities $\Psi$ that is unknown a priori. However, given that $\Psi\leq \rho$ we can safely replace $\Psi$ with $\rho$ in the equation and get the following sufficient sample size:
\begin{align*}
    r\in \bigO\left(\frac{\hat{v}+\varepsilon}{\varepsilon^2}\ln\left(\frac{\rho}{\delta\hat{v}}\right)\right)
\end{align*}
\end{proof}

\section{Missing Pseudocode}\label{apx:pseudocodes}

\begin{algorithm2e}[htb!]
\caption{PercolationDifferences}\label{algo:perc_diff}
\KwIn{Array $A$ sorted in non-decreasing order and $n=|A|$. }
\KwOut{The summation $\sum_{i=1}^{n}\sum_{j=1}^n R(A[j]-A[i]$ and the array $\{\textit{minus}_{\_s}[k] = \sum_{\substack{i = 1\\i\neq k}}\sum_{\substack{j = 1\\ j\neq k}} R(A[j]-A[i]),\forall k\in [1,n]\}$ where $R(x) = \max\{z,0\}$}

$\text{sum}  =0 $

$\text{minus\_s}[i] = 0,\; \forall i\in [1,n]$

$\text{svp}[i] = 0,\; \forall i\in [1,n+1]$

\For{$i = 2$ \textbf{ to }$n$}{
$\text{svp}[i] = \text{svp}[i-1]+A[i-1]$

$\text{sum} = \text{sum}+ (i-1)A[i]-\text{svp}[i] $

}
$\text{svp}[n+1]  = \text{svp}[n]+ A[n]$

\For{$i = 1$\textbf{ to }$n$}{

$\text{minus\_s}[i] = \text{sum}[i]-A[i](2i-n2)-\text{svp}[n+1]+2\text{svp}[i]$

}
\Return $\text{sum},\text{minus\_s}$
\end{algorithm2e}


\newpage
\section{Additional Experiments}\label{apx:experiments}

\begin{table*}[htb!]
	\centering
	\scriptsize
	\begin{tabular}{llllllllllll}
		&                                      & \multicolumn{4}{c}{\textbf{Running times (s)}}                                                                                                                   & \multicolumn{1}{c}{\textbf{}}           & \multicolumn{1}{c}{\textbf{}}                              & \multicolumn{4}{c}{\textbf{Running times (s)}}                                                                                                      \\ \cline{3-12} 
		\textbf{Graph}                                                   & \multicolumn{1}{c}{$\bm{\varepsilon}$} & \multicolumn{1}{c}{\textbf{Algo.~\ref{algo:apx_algo}}}    & \multicolumn{1}{c}{\textbf{Progr. Lima et al.,\cite{Lima_2022}}} & \multicolumn{1}{c}{\textbf{Fixed ss.~\cite{Lima_2020}}} & \multicolumn{1}{c}{\textbf{Exact}}              & \textbf{}                               & \textbf{Graph}                                             & \multicolumn{1}{c}{\textbf{Algo.~\ref{algo:apx_algo}}}    & \multicolumn{1}{c}{\textbf{Progr. Lima et al.,\cite{Lima_2022}}} & \multicolumn{1}{c}{\textbf{Fixed ss.~\cite{Lima_2020}}} & \multicolumn{1}{c}{\textbf{exact}} \\ \hline
		\multicolumn{1}{l|}{\multirow{5}{*}{\rotatebox[origin=c]{90}{\texttt{Musae-FB.}}}} & \multicolumn{1}{l|}{0.1}             & \multicolumn{1}{l|}{\textbf{0.931}}    & \multicolumn{1}{l|}{2.111}       & \multicolumn{1}{l|}{0.936}         & \multicolumn{1}{l|}{\multirow{5}{*}{88.871}}    & \multicolumn{1}{l|}{\multirow{40}{*}{}} & \multicolumn{1}{l|}{\multirow{5}{*}{\rotatebox[origin=c]{90}{\texttt{Web-Notred.}}}}   & \multicolumn{1}{l|}{\textbf{6.523}}    & \multicolumn{1}{l|}{14.828}      & \multicolumn{1}{l|}{12.349}        & \multirow{5}{*}{6478.278}          \\
		\multicolumn{1}{l|}{}                                            & \multicolumn{1}{l|}{0.07}            & \multicolumn{1}{l|}{\textbf{1.341}}    & \multicolumn{1}{l|}{3.896}       & \multicolumn{1}{l|}{1.756}         & \multicolumn{1}{l|}{}                           & \multicolumn{1}{l|}{}                   & \multicolumn{1}{l|}{}                                      & \multicolumn{1}{l|}{\textbf{9.211}}    & \multicolumn{1}{l|}{31.625}      & \multicolumn{1}{l|}{24.789}        &                                    \\
		\multicolumn{1}{l|}{}                                            & \multicolumn{1}{l|}{0.05}            & \multicolumn{1}{l|}{\textbf{2.172}}    & \multicolumn{1}{l|}{8.137}       & \multicolumn{1}{l|}{3.427}         & \multicolumn{1}{l|}{}                           & \multicolumn{1}{l|}{}                   & \multicolumn{1}{l|}{}                                      & \multicolumn{1}{l|}{\textbf{13.025}}   & \multicolumn{1}{l|}{67.775}      & \multicolumn{1}{l|}{51.556}        &                                    \\
		\multicolumn{1}{l|}{}                                            & \multicolumn{1}{l|}{0.01}            & \multicolumn{1}{l|}{\textbf{8.360}}    & \multicolumn{1}{l|}{158.635}     & \multicolumn{1}{l|}{82.225}        & \multicolumn{1}{l|}{}                           & \multicolumn{1}{l|}{}                   & \multicolumn{1}{l|}{}                                      & \multicolumn{1}{l|}{\textbf{123.358}}  & \multicolumn{1}{l|}{1796.095}    & \multicolumn{1}{l|}{1474.979}      &                                    \\
		\multicolumn{1}{l|}{}                                            & \multicolumn{1}{l|}{0.005}           & \multicolumn{1}{l|}{\textbf{19.174}}   & \multicolumn{1}{l|}{567.580}     & \multicolumn{1}{l|}{329.307}       & \multicolumn{1}{l|}{}                           & \multicolumn{1}{l|}{}                   & \multicolumn{1}{l|}{}                                      & \multicolumn{1}{l|}{\textbf{221.356}}  & \multicolumn{1}{l|}{7308.966}    & \multicolumn{1}{l|}{6671.108}      &                                    \\ \cline{1-6} \cline{8-12} 
		\multicolumn{1}{l|}{\multirow{5}{*}{\rotatebox[origin=c]{90}{\texttt{Enron}}}}           & \multicolumn{1}{l|}{0.1}             & \multicolumn{1}{l|}{\textbf{1.552}}    & \multicolumn{1}{l|}{3.117}       & \multicolumn{1}{l|}{1.679}         & \multicolumn{1}{l|}{\multirow{5}{*}{172.607}}   & \multicolumn{1}{l|}{}                   & \multicolumn{1}{l|}{\multirow{5}{*}{\rotatebox[origin=c]{90}{\texttt{Web-Google}}}}      & \multicolumn{1}{l|}{\textbf{18.862}}   & \multicolumn{1}{l|}{80.178}      & \multicolumn{1}{l|}{65.315}        & \multirow{5}{*}{132204.252}        \\
		\multicolumn{1}{l|}{}                                            & \multicolumn{1}{l|}{0.07}            & \multicolumn{1}{l|}{\textbf{2.250}}    & \multicolumn{1}{l|}{5.344}       & \multicolumn{1}{l|}{3.014}         & \multicolumn{1}{l|}{}                           & \multicolumn{1}{l|}{}                   & \multicolumn{1}{l|}{}                                      & \multicolumn{1}{l|}{\textbf{25.749}}   & \multicolumn{1}{l|}{140.140}     & \multicolumn{1}{l|}{130.544}       &                                    \\
		\multicolumn{1}{l|}{}                                            & \multicolumn{1}{l|}{0.05}            & \multicolumn{1}{l|}{\textbf{3.910}}    & \multicolumn{1}{l|}{11.042}      & \multicolumn{1}{l|}{5.740}         & \multicolumn{1}{l|}{}                           & \multicolumn{1}{l|}{}                   & \multicolumn{1}{l|}{}                                      & \multicolumn{1}{l|}{\textbf{44.296}}   & \multicolumn{1}{l|}{286.302}     & \multicolumn{1}{l|}{249.315}       &                                    \\
		\multicolumn{1}{l|}{}                                            & \multicolumn{1}{l|}{0.01}            & \multicolumn{1}{l|}{\textbf{13.845}}   & \multicolumn{1}{l|}{229.725}     & \multicolumn{1}{l|}{139.445}       & \multicolumn{1}{l|}{}                           & \multicolumn{1}{l|}{}                   & \multicolumn{1}{l|}{}                                      & \multicolumn{1}{l|}{\textbf{319.639}}  & \multicolumn{1}{l|}{5751.240}    & \multicolumn{1}{l|}{6048.442}      &                                    \\
		\multicolumn{1}{l|}{}                                            & \multicolumn{1}{l|}{0.005}           & \multicolumn{1}{l|}{\textbf{26.573}}   & \multicolumn{1}{l|}{848.157}     & \multicolumn{1}{l|}{543.379}       & \multicolumn{1}{l|}{}                           & \multicolumn{1}{l|}{}                   & \multicolumn{1}{l|}{}                                      & \multicolumn{1}{l|}{\textbf{642.069}}  & \multicolumn{1}{l|}{22332.602}   & \multicolumn{1}{l|}{24188.226}     &                                    \\ \cline{1-6} \cline{8-12} 
		\multicolumn{1}{l|}{\multirow{5}{*}{\rotatebox[origin=c]{90}{\texttt{CA-Astro.}}}}            & \multicolumn{1}{l|}{0.1}             & \multicolumn{1}{l|}{\textbf{1.173}}    & \multicolumn{1}{l|}{2.091}       & \multicolumn{1}{l|}{1.003}         & \multicolumn{1}{l|}{\multirow{5}{*}{57.195}}    & \multicolumn{1}{l|}{}                   & \multicolumn{1}{l|}{\multirow{5}{*}{\rotatebox[origin=c]{90}{\texttt{Web-Berk.}}}}    & \multicolumn{1}{l|}{\textbf{10.193}}   & \multicolumn{1}{l|}{78.652}      & \multicolumn{1}{l|}{65.751}        & \multirow{5}{*}{93085.472}         \\
		\multicolumn{1}{l|}{}                                            & \multicolumn{1}{l|}{0.07}            & \multicolumn{1}{l|}{\textbf{1.892}}    & \multicolumn{1}{l|}{3.403}       & \multicolumn{1}{l|}{1.913}         & \multicolumn{1}{l|}{}                           & \multicolumn{1}{l|}{}                   & \multicolumn{1}{l|}{}                                      & \multicolumn{1}{l|}{\textbf{11.011}}   & \multicolumn{1}{l|}{137.584}     & \multicolumn{1}{l|}{122.706}       &                                    \\
		\multicolumn{1}{l|}{}                                            & \multicolumn{1}{l|}{0.05}            & \multicolumn{1}{l|}{\textbf{2.806}}    & \multicolumn{1}{l|}{7.144}       & \multicolumn{1}{l|}{3.532}         & \multicolumn{1}{l|}{}                           & \multicolumn{1}{l|}{}                   & \multicolumn{1}{l|}{}                                      & \multicolumn{1}{l|}{\textbf{12.371}}   & \multicolumn{1}{l|}{245.223}     & \multicolumn{1}{l|}{236.836}       &                                    \\
		\multicolumn{1}{l|}{}                                            & \multicolumn{1}{l|}{0.01}            & \multicolumn{1}{l|}{\textbf{9.031}}    & \multicolumn{1}{l|}{139.721}     & \multicolumn{1}{l|}{84.645}        & \multicolumn{1}{l|}{}                           & \multicolumn{1}{l|}{}                   & \multicolumn{1}{l|}{}                                      & \multicolumn{1}{l|}{\textbf{97.808}}   & \multicolumn{1}{l|}{3909.436}    & \multicolumn{1}{l|}{5636.873}      &                                    \\
		\multicolumn{1}{l|}{}                                            & \multicolumn{1}{l|}{0.005}           & \multicolumn{1}{l|}{\textbf{19.519}}   & \multicolumn{1}{l|}{489.967}     & \multicolumn{1}{l|}{314.285}       & \multicolumn{1}{l|}{}                           & \multicolumn{1}{l|}{}                   & \multicolumn{1}{l|}{}                                      & \multicolumn{1}{l|}{\textbf{306.740}}  & \multicolumn{1}{l|}{15055.302}   & \multicolumn{1}{l|}{23562.770}     &                                    \\ \cline{1-6} \cline{8-12} 
		\multicolumn{1}{l|}{\multirow{5}{*}{\rotatebox[origin=c]{90}{\texttt{Epinions}}}}          & \multicolumn{1}{l|}{0.1}             & \multicolumn{1}{l|}{\textbf{2.094}}    & \multicolumn{1}{l|}{3.797}       & \multicolumn{1}{l|}{2.353}         & \multicolumn{1}{l|}{\multirow{5}{*}{508.195}}   & \multicolumn{1}{l|}{}                   & \multicolumn{1}{l|}{\multirow{5}{*}{\rotatebox[origin=c]{90}{\texttt{Flickr}}}}           & \multicolumn{1}{l|}{\textbf{76.865}}   & \multicolumn{1}{l|}{316.055}     & \multicolumn{1}{l|}{352.840}       & \multirow{5}{*}{--}                \\
		\multicolumn{1}{l|}{}                                            & \multicolumn{1}{l|}{0.07}            & \multicolumn{1}{l|}{\textbf{3.089}}    & \multicolumn{1}{l|}{7.465}       & \multicolumn{1}{l|}{5.264}         & \multicolumn{1}{l|}{}                           & \multicolumn{1}{l|}{}                   & \multicolumn{1}{l|}{}                                      & \multicolumn{1}{l|}{\textbf{186.673}}  & \multicolumn{1}{l|}{738.009}     & \multicolumn{1}{l|}{750.455}       &                                    \\
		\multicolumn{1}{l|}{}                                            & \multicolumn{1}{l|}{0.05}            & \multicolumn{1}{l|}{\textbf{5.394}}    & \multicolumn{1}{l|}{16.188}      & \multicolumn{1}{l|}{9.832}         & \multicolumn{1}{l|}{}                           & \multicolumn{1}{l|}{}                   & \multicolumn{1}{l|}{}                                      & \multicolumn{1}{l|}{\textbf{264.211}}  & \multicolumn{1}{l|}{1288.315}    & \multicolumn{1}{l|}{1320.336}      &                                    \\
		\multicolumn{1}{l|}{}                                            & \multicolumn{1}{l|}{0.01}            & \multicolumn{1}{l|}{\textbf{24.815}}   & \multicolumn{1}{l|}{369.019}     & \multicolumn{1}{l|}{237.600}       & \multicolumn{1}{l|}{}                           & \multicolumn{1}{l|}{}                   & \multicolumn{1}{l|}{}                                      & \multicolumn{1}{l|}{\textbf{1169.527}} & \multicolumn{1}{l|}{30233.900}   & \multicolumn{1}{l|}{33421.212}     &                                    \\
		\multicolumn{1}{l|}{}                                            & \multicolumn{1}{l|}{0.005}           & \multicolumn{1}{l|}{\textbf{51.626}}   & \multicolumn{1}{l|}{1403.212}    & \multicolumn{1}{l|}{924.148}       & \multicolumn{1}{l|}{}                           & \multicolumn{1}{l|}{}                   & \multicolumn{1}{l|}{}                                      & \multicolumn{1}{l|}{\textbf{2141.513}} & \multicolumn{1}{l|}{123584.982}  & \multicolumn{1}{l|}{136549.746}    &                                    \\ \cline{1-6} \cline{8-12} 
		\multicolumn{1}{l|}{\multirow{5}{*}{\rotatebox[origin=c]{90}{\texttt{Slashdot}}}}          & \multicolumn{1}{l|}{0.1}             & \multicolumn{1}{l|}{\textbf{2.603}}    & \multicolumn{1}{l|}{6.470}       & \multicolumn{1}{l|}{3.904}         & \multicolumn{1}{l|}{\multirow{5}{*}{1019.431}}  & \multicolumn{1}{l|}{}                   & \multicolumn{1}{l|}{\multirow{5}{*}{\rotatebox[origin=c]{90}{\texttt{Wiki-Talk}}}}       & \multicolumn{1}{l|}{\textbf{32.917}}   & \multicolumn{1}{l|}{145.679}     & \multicolumn{1}{l|}{160.136}       & \multirow{5}{*}{--}                \\
		\multicolumn{1}{l|}{}                                            & \multicolumn{1}{l|}{0.07}            & \multicolumn{1}{l|}{\textbf{4.009}}    & \multicolumn{1}{l|}{11.960}      & \multicolumn{1}{l|}{7.879}         & \multicolumn{1}{l|}{}                           & \multicolumn{1}{l|}{}                   & \multicolumn{1}{l|}{}                                      & \multicolumn{1}{l|}{\textbf{46.465}}   & \multicolumn{1}{l|}{250.890}     & \multicolumn{1}{l|}{253.754}       &                                    \\
		\multicolumn{1}{l|}{}                                            & \multicolumn{1}{l|}{0.05}            & \multicolumn{1}{l|}{\textbf{6.260}}    & \multicolumn{1}{l|}{24.054}      & \multicolumn{1}{l|}{14.859}        & \multicolumn{1}{l|}{}                           & \multicolumn{1}{l|}{}                   & \multicolumn{1}{l|}{}                                      & \multicolumn{1}{l|}{\textbf{80.328}}   & \multicolumn{1}{l|}{428.941}     & \multicolumn{1}{l|}{438.835}       &                                    \\
		\multicolumn{1}{l|}{}                                            & \multicolumn{1}{l|}{0.01}            & \multicolumn{1}{l|}{\textbf{29.466}}   & \multicolumn{1}{l|}{520.693}     & \multicolumn{1}{l|}{346.211}       & \multicolumn{1}{l|}{}                           & \multicolumn{1}{l|}{}                   & \multicolumn{1}{l|}{}                                      & \multicolumn{1}{l|}{\textbf{623.053}}  & \multicolumn{1}{l|}{16788.196}   & \multicolumn{1}{l|}{18587.797}     &                                    \\
		\multicolumn{1}{l|}{}                                            & \multicolumn{1}{l|}{0.005}           & \multicolumn{1}{l|}{\textbf{63.729}}   & \multicolumn{1}{l|}{1954.067}    & \multicolumn{1}{l|}{1333.640}      & \multicolumn{1}{l|}{}                           & \multicolumn{1}{l|}{}                   & \multicolumn{1}{l|}{}                                      & \multicolumn{1}{l|}{\textbf{1329.545}} & \multicolumn{1}{l|}{76922.028}   & \multicolumn{1}{l|}{85060.659}     &                                    \\ \cline{1-6} \cline{8-12} 
		\multicolumn{1}{l|}{\multirow{5}{*}{\rotatebox[origin=c]{90}{\texttt{Cit-HepPh}}}}             & \multicolumn{1}{l|}{0.1}             & \multicolumn{1}{l|}{\textbf{1.043}}    & \multicolumn{1}{l|}{2.683}       & \multicolumn{1}{l|}{1.563}         & \multicolumn{1}{l|}{\multirow{5}{*}{99.647}}    & \multicolumn{1}{l|}{}                   & \multicolumn{1}{l|}{\multirow{5}{*}{\rotatebox[origin=c]{90}{\texttt{Twitter}}}} & \multicolumn{1}{l|}{\textbf{51.957}}   & \multicolumn{1}{l|}{234.506}     & \multicolumn{1}{l|}{253.671}       & \multirow{5}{*}{--}                \\
		\multicolumn{1}{l|}{}                                            & \multicolumn{1}{l|}{0.07}            & \multicolumn{1}{l|}{\textbf{1.257}}    & \multicolumn{1}{l|}{4.872}       & \multicolumn{1}{l|}{2.316}         & \multicolumn{1}{l|}{}                           & \multicolumn{1}{l|}{}                   & \multicolumn{1}{l|}{}                                      & \multicolumn{1}{l|}{\textbf{71.298}}   & \multicolumn{1}{l|}{381.467}     & \multicolumn{1}{l|}{392.843}       &                                    \\
		\multicolumn{1}{l|}{}                                            & \multicolumn{1}{l|}{0.05}            & \multicolumn{1}{l|}{\textbf{1.877}}    & \multicolumn{1}{l|}{8.322}       & \multicolumn{1}{l|}{5.157}         & \multicolumn{1}{l|}{}                           & \multicolumn{1}{l|}{}                   & \multicolumn{1}{l|}{}                                      & \multicolumn{1}{l|}{\textbf{101.319}}  & \multicolumn{1}{l|}{540.377}     & \multicolumn{1}{l|}{550.069}       &                                    \\
		\multicolumn{1}{l|}{}                                            & \multicolumn{1}{l|}{0.01}            & \multicolumn{1}{l|}{\textbf{11.938}}   & \multicolumn{1}{l|}{152.188}     & \multicolumn{1}{l|}{121.365}       & \multicolumn{1}{l|}{}                           & \multicolumn{1}{l|}{}                   & \multicolumn{1}{l|}{}                                      & \multicolumn{1}{l|}{\textbf{815.155}}  & \multicolumn{1}{l|}{22014.327}   & \multicolumn{1}{l|}{24382.831}     &                                    \\
		\multicolumn{1}{l|}{}                                            & \multicolumn{1}{l|}{0.005}           & \multicolumn{1}{l|}{\textbf{23.858}}   & \multicolumn{1}{l|}{580.238}     & \multicolumn{1}{l|}{518.887}       & \multicolumn{1}{l|}{}                           & \multicolumn{1}{l|}{}                   & \multicolumn{1}{l|}{}                                      & \multicolumn{1}{l|}{\textbf{1599.591}} & \multicolumn{1}{l|}{92507.598}   & \multicolumn{1}{l|}{102296.969}    &                                    \\ \cline{1-6} \cline{8-12} 
		\multicolumn{1}{l|}{\multirow{5}{*}{\rotatebox[origin=c]{90}{\texttt{Twitch}}}}   & \multicolumn{1}{l|}{0.1}             & \multicolumn{1}{l|}{\textbf{16.622}}   & \multicolumn{1}{l|}{42.095}      & \multicolumn{1}{l|}{24.612}        & \multicolumn{1}{l|}{\multirow{5}{*}{19850.168}} & \multicolumn{1}{l|}{}                   & \multicolumn{1}{l|}{\multirow{5}{*}{\rotatebox[origin=c]{90}{\texttt{soc-Pokec}}}}       & \multicolumn{1}{l|}{\textbf{32.599}}   & \multicolumn{1}{l|}{144.823}     & \multicolumn{1}{l|}{159.277}       & \multirow{5}{*}{--}                \\
		\multicolumn{1}{l|}{}                                            & \multicolumn{1}{l|}{0.07}            & \multicolumn{1}{l|}{\textbf{30.810}}   & \multicolumn{1}{l|}{71.168}      & \multicolumn{1}{l|}{42.678}        & \multicolumn{1}{l|}{}                           & \multicolumn{1}{l|}{}                   & \multicolumn{1}{l|}{}                                      & \multicolumn{1}{l|}{\textbf{50.363}}   & \multicolumn{1}{l|}{267.129}     & \multicolumn{1}{l|}{269.391}       &                                    \\
		\multicolumn{1}{l|}{}                                            & \multicolumn{1}{l|}{0.05}            & \multicolumn{1}{l|}{\textbf{60.612}}   & \multicolumn{1}{l|}{121.481}     & \multicolumn{1}{l|}{77.830}        & \multicolumn{1}{l|}{}                           & \multicolumn{1}{l|}{}                   & \multicolumn{1}{l|}{}                                      & \multicolumn{1}{l|}{\textbf{85.940}}   & \multicolumn{1}{l|}{459.032}     & \multicolumn{1}{l|}{468.766}       &                                    \\
		\multicolumn{1}{l|}{}                                            & \multicolumn{1}{l|}{0.01}            & \multicolumn{1}{l|}{\textbf{1038.137}} & \multicolumn{1}{l|}{2586.024}    & \multicolumn{1}{l|}{1869.815}      & \multicolumn{1}{l|}{}                           & \multicolumn{1}{l|}{}                   & \multicolumn{1}{l|}{}                                      & \multicolumn{1}{l|}{\textbf{443.492}}  & \multicolumn{1}{l|}{11911.870}   & \multicolumn{1}{l|}{13184.466}     &                                    \\
		\multicolumn{1}{l|}{}                                            & \multicolumn{1}{l|}{0.005}           & \multicolumn{1}{l|}{\textbf{1592.149}} & \multicolumn{1}{l|}{10655.417}   & \multicolumn{1}{l|}{7404.675}      & \multicolumn{1}{l|}{}                           & \multicolumn{1}{l|}{}                   & \multicolumn{1}{l|}{}                                      & \multicolumn{1}{l|}{\textbf{1071.343}} & \multicolumn{1}{l|}{62021.929}   & \multicolumn{1}{l|}{68590.269}     &                                    \\ \cline{1-6} \cline{8-12} 
		\multicolumn{1}{l|}{\multirow{5}{*}{\rotatebox[origin=c]{90}{\texttt{Gnutella31}}}}        & \multicolumn{1}{l|}{0.1}             & \multicolumn{1}{l|}{\textbf{1.499}}    & \multicolumn{1}{l|}{3.064}       & \multicolumn{1}{l|}{2.244}         & \multicolumn{1}{l|}{\multirow{5}{*}{237.270}}   & \multicolumn{1}{l|}{}                   &                                                            &                                        &                                  &                                    &                                    \\
		\multicolumn{1}{l|}{}                                            & \multicolumn{1}{l|}{0.07}            & \multicolumn{1}{l|}{\textbf{2.092}}    & \multicolumn{1}{l|}{5.308}       & \multicolumn{1}{l|}{4.669}         & \multicolumn{1}{l|}{}                           & \multicolumn{1}{l|}{}                   &                                                            &                                        &                                  &                                    &                                    \\
		\multicolumn{1}{l|}{}                                            & \multicolumn{1}{l|}{0.05}            & \multicolumn{1}{l|}{\textbf{3.413}}    & \multicolumn{1}{l|}{12.555}      & \multicolumn{1}{l|}{7.989}         & \multicolumn{1}{l|}{}                           & \multicolumn{1}{l|}{}                   &                                                            &                                        &                                  &                                    &                                    \\
		\multicolumn{1}{l|}{}                                            & \multicolumn{1}{l|}{0.01}            & \multicolumn{1}{l|}{\textbf{18.184}}   & \multicolumn{1}{l|}{237.955}     & \multicolumn{1}{l|}{186.660}       & \multicolumn{1}{l|}{}                           & \multicolumn{1}{l|}{}                   &                                                            &                                        &                                  &                                    &                                    \\
		\multicolumn{1}{l|}{}                                            & \multicolumn{1}{l|}{0.005}           & \multicolumn{1}{l|}{\textbf{34.947}}   & \multicolumn{1}{l|}{903.453}     & \multicolumn{1}{l|}{725.400}       & \multicolumn{1}{l|}{}                           & \multicolumn{1}{l|}{}                   &                                                            &                                        &                                  &                                    &                                    \\ \cline{1-7}
	\end{tabular}
\caption{Running times (in seconds) of all the considered algorithms.}
\label{tab:running_times}
\end{table*}

\end{document}